\documentclass[a4paper,pre,twocolumn,showpacs, showkeys,superscriptaddress]{revtex4}
\usepackage[english]{babel}
\usepackage[utf8]{inputenc}
\usepackage{graphicx,epsfig,placeins}
\usepackage{latexsym}
\usepackage{hyperref}
\usepackage{amsfonts} 
\usepackage{amsmath}
\usepackage{textcomp}
\usepackage{epstopdf}
\usepackage{color}
\usepackage{appendix}
\usepackage{stackengine}

\def\ket|#1>{| #1 \rangle}
\def\bra<#1|{\langle #1 |}
\def\<{\left\langle}
\def\>{\right\rangle}
\def\({\left(}
\def\){\right)}
\def\[{\left[}
\def\]{\right]}
\def\{{\left\lbrace}
\def\}{\right\rbrace}
\def\beq{\begin{equation}}
\def\eeq{\end{equation}}

\def\N{{\mathbb N}}
\def\R{{\mathbb R}}
\def\e{{\mathrm e}}

\begin{document}

\title{Synchronization of fluctuating delay-coupled chaotic networks}%

\author{Manuel Jiménez}
\affiliation{Departamento de Física Fundamental, UNED, Spain}%
\author{Javier Rodríguez-Laguna} 
\affiliation{Departamento de Física Fundamental, UNED, Spain}%
\author{Otti D'Huys} 
\affiliation{Department of Mathematics, Aston University, B7 4ET Birmingham, United Kingdom}%
\author{Javier de la Rubia}
\affiliation{Departamento de Física Fundamental, UNED, Spain}%
\author{Elka Korutcheva}
\affiliation{Departamento de Física Fundamental, UNED, Spain}%
\affiliation{G. Nadjakov Inst. Solid State Physics, Bulgarian Academy
  of Sciences, 1784, Sofia, Bulgaria} 

\begin{abstract}
We study the synchronization of chaotic units connected through time-delayed fluctuating interactions. We focus on small-world networks of Bernoulli and Logistic units with a fixed chiral backbone. Comparing the synchronization properties of static and fluctuating networks, we find that random network alternations can enhance the synchronizability. Synchronized states appear to be maximally stable when fluctuations are much faster
than the time-delay, even when the instantaneous state of the network does not allow synchronization. This enhancing effect disappears for very slow fluctuations.  For fluctuation time scales of the order of the time-delay, a desynchronizing resonance is reported. Moreover, we observe characteristic oscillations, with a periodicity related to the time-delay, as the system approaches or drifts away from the synchronized state.
\end{abstract}


\date{\today}

\maketitle

\section{Introduction}

Cooperative behavior of chaotic systems in interaction can lead to the emergence of partial and local synchronization \cite{Boccaletti2002}. An interesting problem in this context is the stability of the synchronized state, which is ruled by the topology of the interaction \cite{Pecora1998,Scholl2010}. In most settings, the coupling terms carry a finite time-delay due to the finite velocity of transmission of information. Yet, even for infinitely large time-delay the units can achieve zero-lag synchronization \cite{Atay2004}.  The paradigmatic time-delayed coupled systems capable of chaos synchronization are semiconductor lasers \cite{Heiligenthal2001, Fischer06, Heiligenthal2011, Nixon2012}, with interesting applications in secure communication \cite{Shore2005, KanterKinzel2008}. The phenomenon might have relevance as well in neuroscience \cite{Buzsaki2009,KanterCohen2011}.

A better understanding of chaos synchronization can be gained by studying simple chaotic systems, such as Bernoulli maps under single \cite{KanterKinzel2011} or multiple \cite{self_citation} time-delays, for
which the conditions for a stable synchronized state can be obtained
analytically. These studies generally perform a stability analysis of
the synchronized state on a fixed interaction network. Additionally, a general formalism has been developed for
ensembles of static random interaction networks \cite{FengMingzhouC2006}. 

Currently, there is an increasing interest in studying networks as time-varying entities \cite{HolmeEPJB2015}. In fact, network fluctuations are essential features of some systems such as, for instance, interacting neurons, where synaptic plasticity continuously changes the topology \cite{BuonomanoMerzenich1998}. It is interesting then to inquire how does a fluctuating network affect synchronization stability \cite{BelykhHasler2004}.
Recently, there has been a number of studies concerning synchronization on
time-varying contact networks, where the topology changes due to the
random motion of the agents, and the couplings are instantaneous. Most
of them consider diffusive coupling of moving oscillators
\cite{PeruaniMorelli2010, FujiwaraGuilera2011}, but also chaotic units
\cite{FrascaBoccaletti2008, FujiwaraGuilera2016}. This problem has also been tackled
for genetic oscillators moving on lattices \cite{UriuMorelli2013, UriuMorelli2014}.
In this paper we study the synchronization properties of chaotic maps
interacting on random small-world directed networks whose topology
fluctuates in time and whose links bear a large time-delay. There has
been a very recent work concerning the case of coupled chaotic maps
interacting with small time-delays comparable to the time scale of
network switching \cite{NagPoriaCS&F2016}. Our work builds on the
previous studies by exploring synchronization stability on the full
range of possible scalings between the time-delay and the network
switching time scale.

We consider an interaction network of coupled chaotic maps with a single  coupling delay, $T_d$. The networks fluctuates with a characteristic time-scale $T_n$. These network fluctuations are random, and not adaptive, i.e., the network evolution is not linked to the state in any way. If $T_n\gg T_d$, i.e., the slow network regime, the dynamics always have enough time to adapt to the current network. Therefore, if the network acquires a desynchronizing configuration, it will lose synchronization. Re-synchronization, although possible in principle, is unlikely, for the chaotic maps that we consider. Thus, we may regard the long term dynamics as desynchronized whenever the probability of reaching a non-synchronizing network is finite. The case where $T_n$ is comparable to $T_d$ or smaller is more involved. Indeed, the system will spend time both in synchronizing and de-synchronizing networks. When the network reaches a de-synchronizing configuration, it will start to escape the synchronization manifold. But, if $T_n$ is
not large enough, there is a probability of returning to a
synchronizing configuration before a certain {\em irreversibility
  line} is crossed. Thus, the system may stay synchronized. Indeed,
our results will prove stronger: when $T_n\ll T_d$, in the fast
fluctuations regime, synchronization becomes significantly more
stable. This is in qualitative agreement with the fast switching approximation
\cite{StilwayRobertson2006}, which states that synchronization is
possible for fast fluctuating networks and diffusive coupling if the
time averaged graph laplacian synchronizes. 


This paper is organized as follows. Section \ref{sec:system} defines
our time-delayed dynamical system, and reviews the basic framework for
its study. In section \ref{sec:sw} we study in detail the
synchronizability in the case of (static) small-world networks. Then,
we define our fluctuating networks, the observables to be employed,
and discuss the numerical results in detail in
\ref{sec:fluctuating}. The last section is devoted to the conclusions
and further work.


\section{Synchronization of delayed chaotic networks}
\label{sec:system}

Let us consider $N$ classical units, characterized by a single degree
of freedom $u_i(t)$, $i\in \{1,\cdots,N\}$ and time $t\in \N$, whose
evolution is given by:

\begin{equation}
u_i(t+1)=(1-\epsilon)f(u_i(t))+\epsilon \sum_j G_{ij}(t)
f(u_j(t-T_d)),
\label{eq:motion}
\end{equation}
where $\epsilon\in[0,1]$ is a real parameter which measures the strength of the interaction, $T_d$ is the coupling delay and $f:[0,1]\mapsto[0,1]$ is a chaotic map. We consider the the Bernoulli map, modeled as

\begin{equation}
f(x)=ax \mod 1,
\label{eq:bernoulli}
\end{equation}
with $a\in\R^+$, and the Logistic map, given by

\begin{equation}
f(x)=rx(1-x),
\label{eq:Logistic}
\end{equation}
with $r\in\R^+$. The network structure is described by the adjacency matrix $G$. To ensure the {\em existence} of a synchronized dynamical state, the adjacency matrix is subject to a unit row-sum condition (also known as stochasticity condition),

\begin{equation}
\sum_j G_{ij}(t)=1,
\label{eq:rowsum}
\end{equation}
for all $i$ and all times $t$. Nonetheless, this does not inform us about the stability of such synchronized state. 

As a measure of the (zero-lag) synchronization in the network, we have chosen the logarithm of the spatial deviation over the network nodes. Let us explain in detail the meaning of
this observable. Consider the spatial average of the unit states
for a given time as

\begin{equation}
\mu(t) \equiv {1\over N} \sum_{i=1}^N u_i(t),
\label{eq:spatial_average}
\end{equation} 
and the corresponding spatial standard deviation as

\begin{equation}
\sigma \equiv \sqrt{ {1\over N}\sum_{i=1}^N (u_i-\mu)^2}.
\label{eq:spatial_deviation}
\end{equation}
Then, the {\em synchronization level} is defined as 

\begin{equation}
{\cal S}\equiv -\ln(\sigma).
\label{eq:sync_level}
\end{equation}

A perfectly synchronized state would have ${\cal S}\to+\infty$, but in practice this value is bounded by the machine precision. In our calculations, using double precision floating point numbers, the maximal synchronization level corresponds approximately to ${\cal S}\sim 35$, which implies a deviation of order $\sigma \sim \exp{[-35]}\sim 10^{-15}$. On the other extreme, in a desynchronized state each unit behaves independently and ${\cal S} = O(1)$. The minimal computed value of ${\cal S}$ is close to $1.95$, which is close to the mean deviation for a uniform distribution on $[0,1]$: $-\ln \sigma = -\ln(12)/2 \approx 1.24$.

We initialize the network close the synchronized state:  all units evolve in unison for $T_d$ time steps, and we apply at $t=0$ a random point-like perturbation $\boldsymbol{\mathbf{\xi}}=A\cdot \boldsymbol{\mathbf{r}}$, with $A = 10^{-10}$ and $\boldsymbol{\mathbf{r}}$ is a vector of random numbers drawn uniformly from $[0,1]$.  Thus, the synchronization level at $t=0$ is always around 25, instead of the machine precision value of 35. We will denote by $\<\cdot\>$ the {\em realization average} over such initial conditions. Our most relevant observable, therefore, will be $\<{\cal S}(t)\>$.

We define a {\em synchronization Lyapunov exponent} (SLE) as the
average linear rate at which the synchronization level increases or decreases with time:

\begin{equation}
\<{\cal S}(t)\> \sim {\cal S}_0 - \lambda t.
\label{eq:sle}
\end{equation}

By this definition, $\lambda$ is equivalent to the maximal Lyapunov exponent transverse to the synchronization manifold, given by the master stability function \cite{Pecora1998}, which approximates the evolution of a perturbation from the synchronized state $\sigma \sim (u_i-\mu)\propto e^{\lambda t}$. 

The stability of the synchronized state is related to the second largest eigenvalue of the adjacency matrix $G$. Let $\{\gamma_i\}_{i=1}^N$ be the eigenvalues of $G$ sorted in descending order of their modulus, $|\gamma_1|\geq |\gamma_2| \geq \cdots \geq |\gamma_N|$. Gerschgorin circle theorem \cite{GolubVanLoan1996} can be applied, showing that $|\gamma_i|\leq 1$, and the unit row sum guarantees that $|\gamma_1|=1$, with eigenvector $v_1=[1, \dots, 1]$. Hence, a perturbation along this mode preserves synchronization as it affects every unit equally. The evolution of a perturbation away from the synchronization manifold will then evolve according to the mode with second largest eigenvalue \cite{Pecora1998}.

For a network of Bernoulli maps (Eq. \eqref{eq:bernoulli}), assuming that $(a(1-\epsilon))^{T_d}\ll1$ and $(a\epsilon|\gamma_2|)^{T_d}\ll 1$ hold, the SLE can be approximated by the following expression \cite{KanterKinzel2008, DHuysKinzel2013}.
\begin{equation}
\lambda\approx \frac{1}{T_d}\ln\left|\frac{a\epsilon\gamma_2}{1-a(1-\epsilon)}\right|
\label{eq:lyap_bernoulli}
\end{equation}
Hence, the condition for a stable synchronization manifold reads:
\begin{equation}
\epsilon> {a-1\over a\Delta}\,,
\label{eq:sync_bernoulli}
\end{equation}
where $\Delta=1-|\gamma_2|$ is the \emph{spectral gap} or \emph{eigengap}.

For other chaotic maps, the stretch factor of the map $|f'(x)|$ is not constant, and the SLE is not analytically accessible. However, it has been shown that fluctuations in the term $|f'(x)|$ along the chaotic trajectory result in a larger spectrum of Lyapunov exponents and thus in a smaller parameter region that sustains stable synchronization \cite{Jungling2015}. We will study synchronization stability on fluctuating networks of coupled Bernouilli maps in order to be able to compare with the analytical results for static networks. We will also study networks of Logistic maps in order to assess the generality of our results.


\section{Synchronization of small-world static networks}
\label{sec:sw}

We have studied the stability of the synchronization manifold on statistical ensembles of
small world networks, which constitute a standard benchmark for network synchronization \cite{BarahonaPecora2002,BelykhHasler2004,Aviad2012,GrabowJurgenPRE2015}.
Here, we will consider a family of Newman-Watts networks \cite{Newman1999}, similar to the
standard small-world \cite{WattsStrogatz1998} but keeping the outside
ring fixed, so that it is guaranteed that the network is always
connected. We will refer generically to these networks as small world (SW).

\begin{figure}
\centering
\epsfig{file=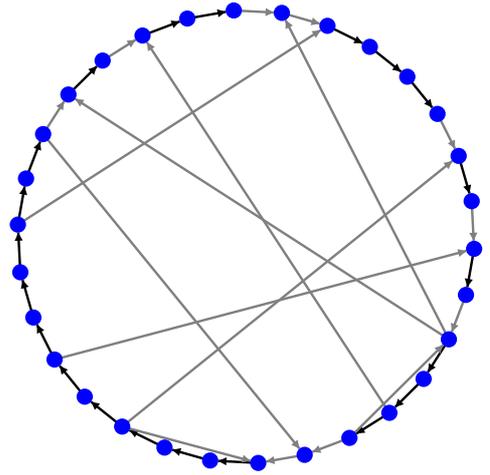,width=7.5cm}
\centering
\epsfig{file=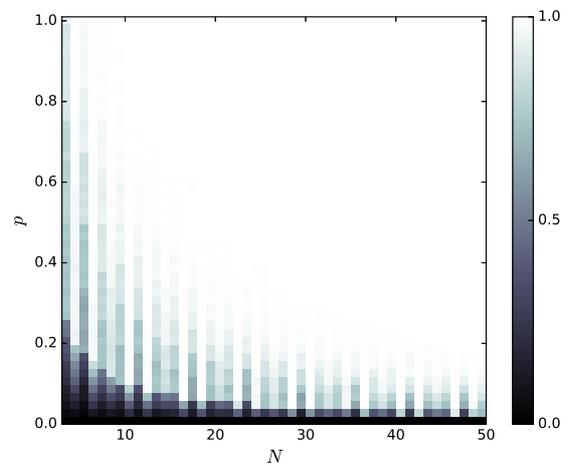, width=9cm}
\caption{(Color online) Top: Illustrating the directed networks used in our dynamical
  systems. In the example, a $N=30$ network with a clockwise rotating
  backbone and $p=0.3$, so the number of shortcuts is $N_s=9$. The
  strength of each link is denoted by its color: black is $1$ and gray
  is $1/2$. Notice that the sum of input links on any node is always
  $1$, as imposed in Eq. \eqref{eq:rowsum}. 
  Bottom: Average fraction of the networks having $GCD=1$ in the $N\times p$
  space. For high enough number of units and shortcuts the probability of
  $GCD >1$ is negligible.}
\label{fig:sw}
\end{figure}

To construct our networks, we consider a chiral 1D chain of $N$ sites, where the only non-zero
entries have the form $G_{i,i+1}$, with periodic boundary conditions. Then we add to the chiral backbone a number of $\langle pN \rangle$ of directed {\em shortcuts}, connecting random sites, with $p\in [0,1]$. An example of such a SW network, with $N=30$ and $p=0.3$ is shown in Fig. \ref{fig:sw} (top) for illustration.

A first requirement for complete synchronization is given by the GCD
condition, which states that the number of possible synchronized
subnetworks is equal to the Greatest Common Divisor (GCD) of the loop
lengths of the network \cite{KanterKinzel2011}. Hence, complete
synchronization is only possible if the GCD of the lengths of all
cycles in the network is unity. This is almost always the case for
large enough networks with a finite number of shortcuts, as can be
seen in Fig. \ref{fig:sw} (bottom). When the GCD condition is met, the
stability of the synchronization manifold is still determined by the
eigengap, as stated in Eq. \eqref{eq:sync_bernoulli} for the specific case of Bernouilli maps.

\begin{figure}
\epsfig{file=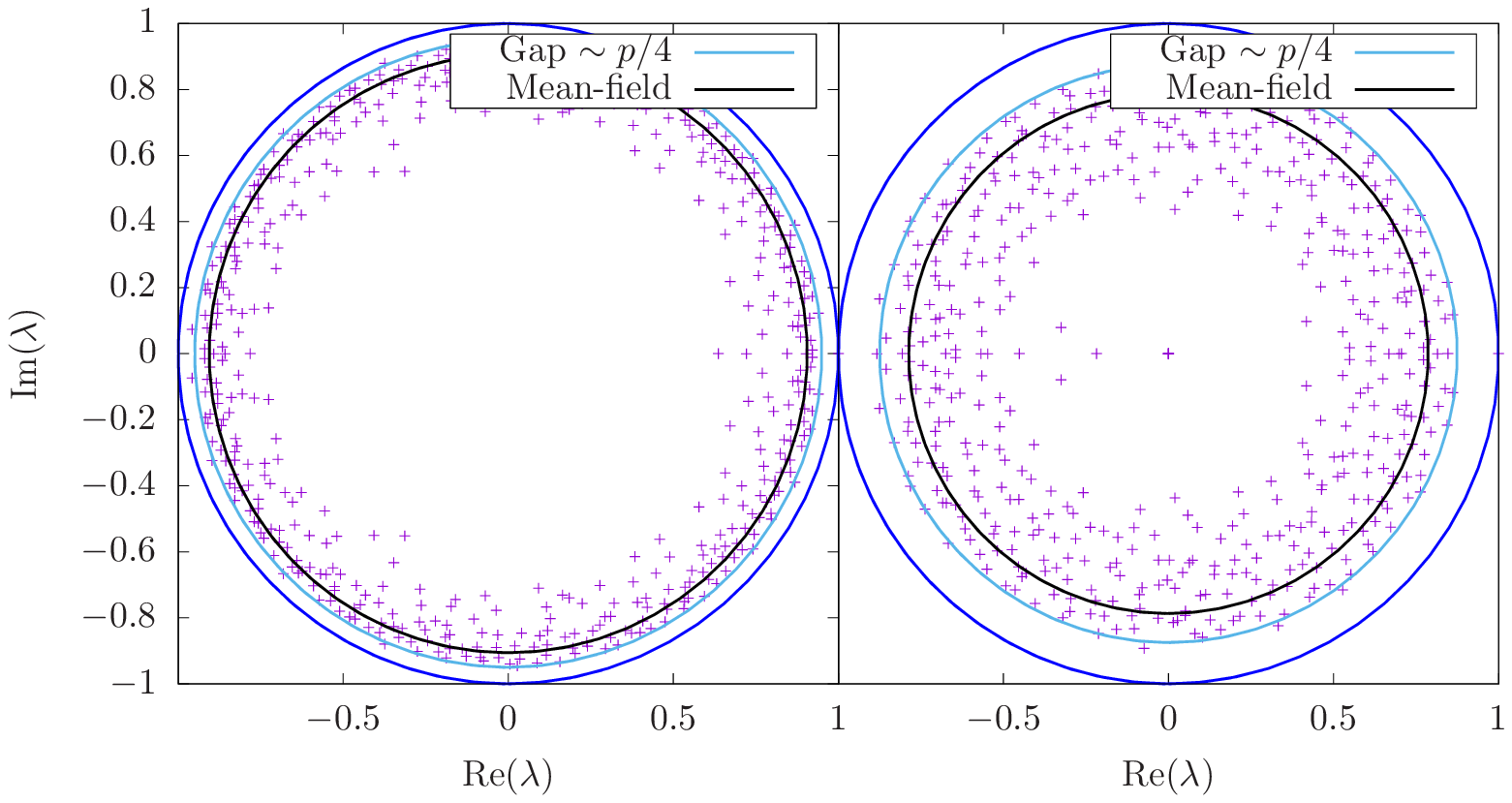,width=9cm}
\epsfig{file=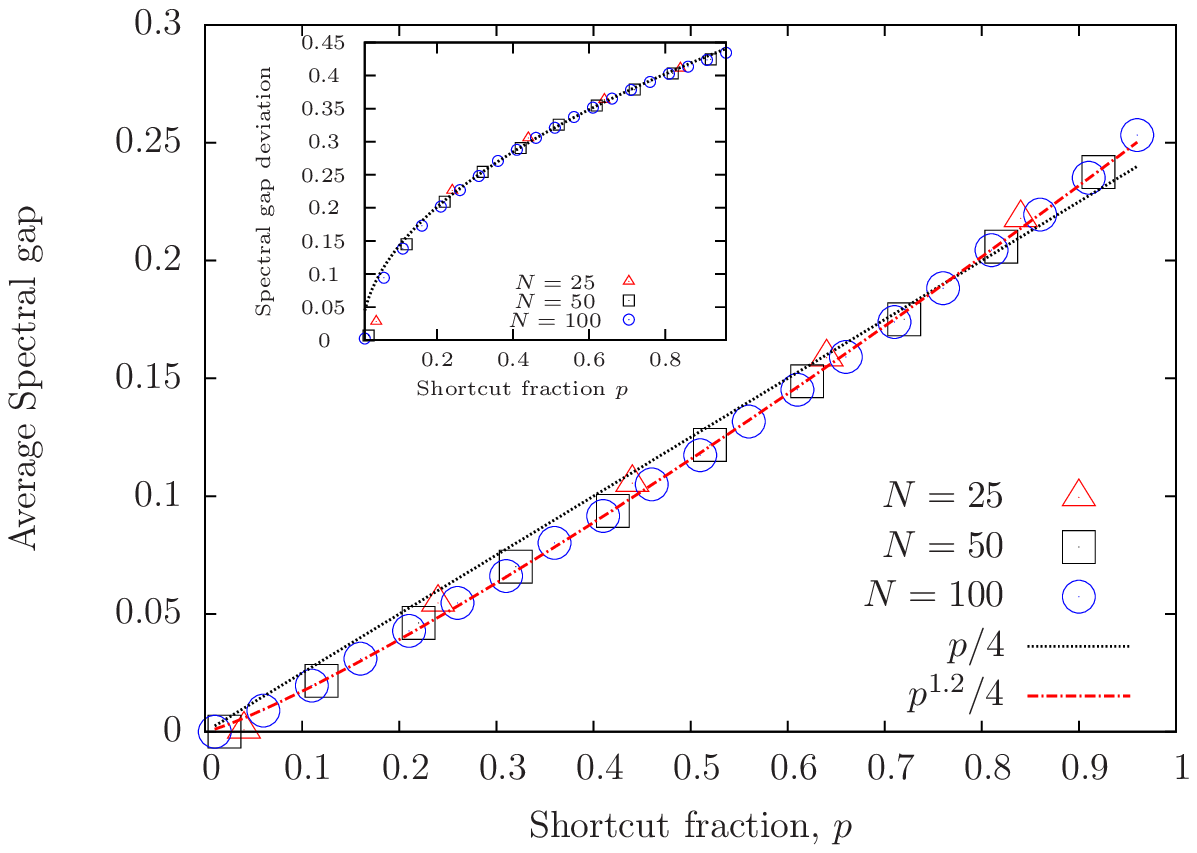,width=8cm}
\caption{(Color online) Top: Spectrum of the adjacency matrix of two SW networks,
  using $N=500$, and $p=1/5$ (left) and $1/2$ (right). Most of the eigenvalues are contained within a ring whose outer radius scales as $p/4$. Bottom: spectral gap for
  different values of $N$ as a function of $p$ of our SW networks, which grows approximately as $p/4$. A more accurate fit, with exponent $1.2$ is also shown. Inset: Standard deviation of the gap scales as the square root of $p$.}
\label{fig:sw_spectrum}
\end{figure}

The adjacency matrices $G$ of directed networks are not hermitian, and
their spectrum need not be real. Let us discuss the statistical
properties of their spectra, in similarity to the studies of
\cite{KuhnJPA2008,GrabowJurgenPRE2015}. Fig. \ref{fig:sw_spectrum}
(top) shows the eigenvalues $\{\gamma_i\}$ on the complex plane for
two SW networks, using $N=500$ and $p=1/5$ (left) and $1/2$
(right). Notice that, following Gerschgorin theorem, they are always
contained within the unit circle. Except for the $\gamma_1=1$
eigenvalue, which is a consequence of the row-sum condition, the
phases of the eigenvalues seem to be homogeneously
distributed. Moreover, they seem to be contained within a ring, whose
radius we would like to characterize.

An interesting approach to estimate the properties of the spectrum of
random matrices describing SW networks was developed in
\cite{GrabowJurgenPRE2015}, using a mean-field approach: write down
the {\em average matrix}, whose entries are given by the average of
the matrix entries. Due to translation invariance, the resulting
matrix is a {\em circulant matrix}, whose spectrum can be found
analytically. We have followed this approach in order to find the mean
field spectrum of our SW networks. In the appendix \ref{ap:MF} we
show that the spectrum of this average matrix lies in the vicinity of
a circumference of radius

\begin{equation}
|\gamma_m^{\textsc{mf}}| \approx  \frac{1-\e^{-p}}{p},
\label{eq:meanfieldresult}
\end{equation}
which appears marked in Fig. \ref{fig:sw_spectrum} (top) as {\em
  mean-field} line. However, the mean-field theory does not describe accurately the ring structure of the spectrum. We have found
numerically that the outer circumference has an approximate radius of
$(1-p/4)$, independent of $N$. Thus, our estimate for the gap is

\begin{equation}
  \Delta \approx p/4\,.
  \label{eq:estimate_gap}
\end{equation}

The lower panel of Fig. \ref{fig:sw_spectrum} shows the numerical
evidence for expression \eqref{eq:estimate_gap}, plotting the average
spectral gap as a function of $p$ for different system sizes
$N$. Interestingly, the inset shows that the standard deviation of the spectral
gap among samples only grows like the square root of $p$,
$\sigma_\Delta \sim p^{1/2}/2$.

Once the spectral properties of our networks have been elucidated we
can proceed to study their dynamics. We have simulated the dynamical system Eq. \eqref{eq:motion} and obtained numerically the
synchronization Lyapunov exponent (SLE) for $10^5$ different SW
networks with $N=40$, $T_d=100$, $\epsilon=0.7$, two values of
$p=0.5$ and $0.8$, and two maps: Bernoulli, Eq. \eqref{eq:bernoulli},
and Logistic, Eq. \eqref{eq:Logistic}. We have chosen the numerical
values of the parameters such that $a=1.1$ and $r=3.577$ so both maps have
comparable Lyapunov exponents when considered in isolation,
\footnote{ 
The Lyapunov exponent of an isolated Bernouilli map is $\lambda_{\textsc{b}}=\ln a\approx 0.09531$. The Lyapunov exponent of the Logistic map, $\lambda_{\textsc{l}}$, must be obtained computationally, but a value of $r=3.577$ gives $\lambda_{\textsc{b}}\approx\lambda_{\textsc{l}}$
\cite{elaydi2007discrete}. }.

\begin{figure}
\epsfig{file=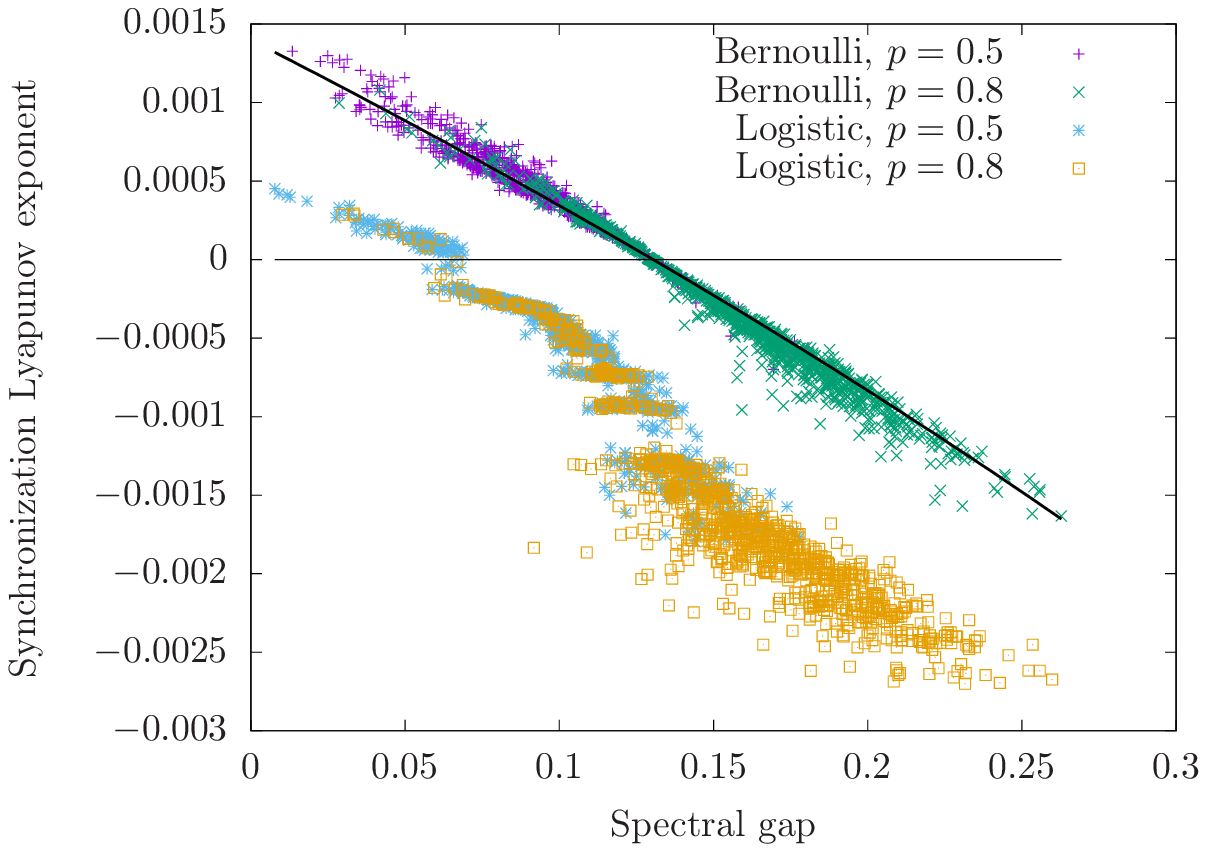,width=8cm}
\epsfig{file=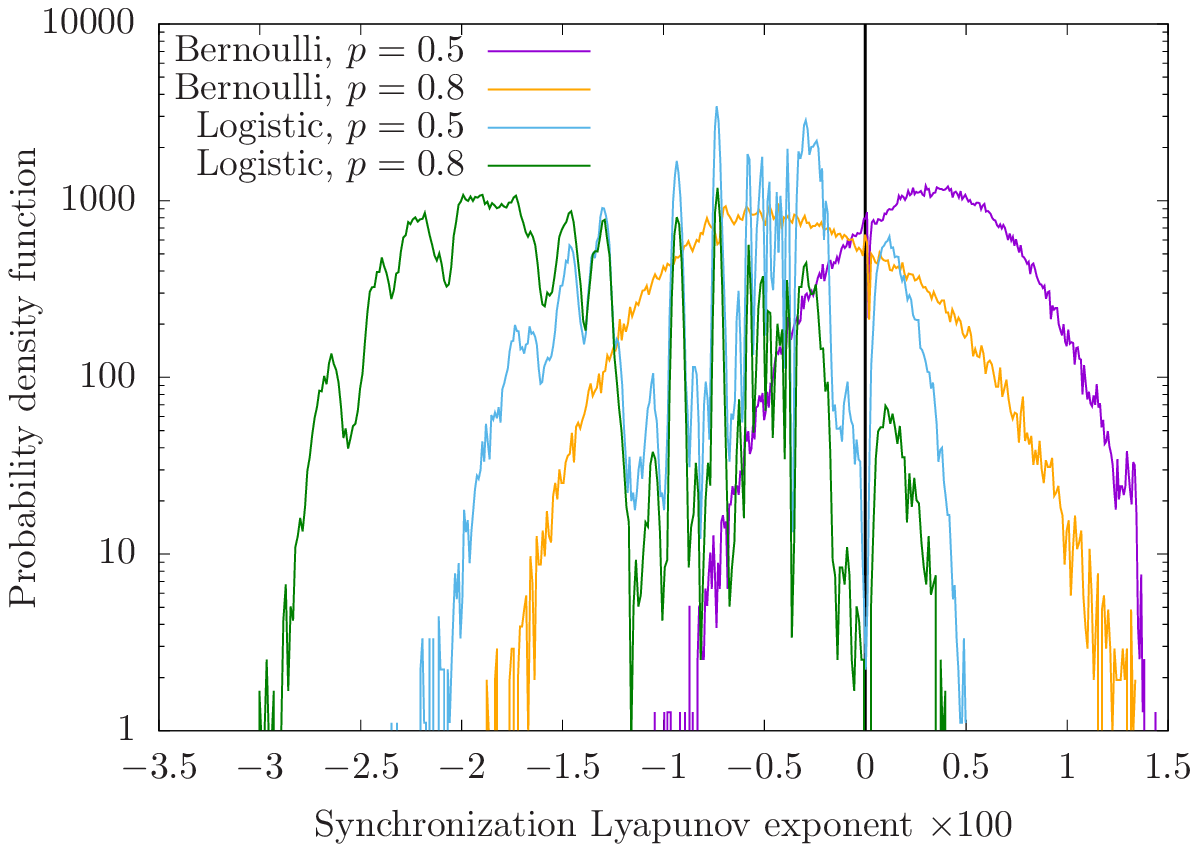,width=8cm}
\caption{(Color online) Measurement of the synchronization Lyapunov exponent (SLE)
  along with the spectral gap for $10^5$ samples of two SW network ensembles with $N=40$ sites,
  $\epsilon=0.7$, and $p=0.5$ and $p=0.8$, for two different
  dynamical systems: Bernoulli and Logistic.  
  Top: Relation between the SLE and the spectral gap. The black
  line corresponds to the theoretical expression Eq. \eqref{eq:lyap_bernoulli} for coupled Bernouilli maps
  with long delays, and captures correctly the measured SLE, specially at values close to zero.
  For the Logistic maps, the correlation is still strong, but much more
  involved. Bottom: SLE histogram for the same four cases, in
  log-scale. Notice that, for Bernoulli we obtain an approximately
  Gaussian behavior, with non-zero skewness. For the Logistic case,
  the SLE are distributed in a much more complicated way. The vertical
  black bar marks the zero SLE, so on the left we have
  synchronization.}
\label{fig:sw_sync}
\end{figure}

In Fig. \ref{fig:sw_sync} (top) we plot the values of the SLE against the spectral gap of each network. In the Bernoulli case we find a tight relationship which follows approximately the theoretical expression for long delays Eq. \eqref{eq:lyap_bernoulli}. The Logistic case is more involved, but the negative correlation between both magnitudes is still clear. We can see that Logistic systems synchronize better than Bernouilli ones as they show a higher number of negative SLE values. Generally, networks with a larger spectral gap synchronize better. Also, there is a higher probability of synchronization for higher $p$ values, since these networks present a larger eigengap.

Fig. \ref{fig:sw_sync} (bottom) shows the histogram of the SLE for
Bernoulli and Logistic systems, using always $N=40$, $T_d=100$,
$\epsilon=0.7$ and two values of $p=0.5$ and $0.8$. The black vertical
bar marks the zero value: points on its left correspond to networks in which the syncronized state is stable. The histograms for the Bernoulli case have a nearly
Gaussian shape, but with finite skewness and kurtosis
\cite{BillenRabinovitchPRE2009}. The histograms are much more involved
for the Logistic case.

\begin{figure}
\epsfig{file=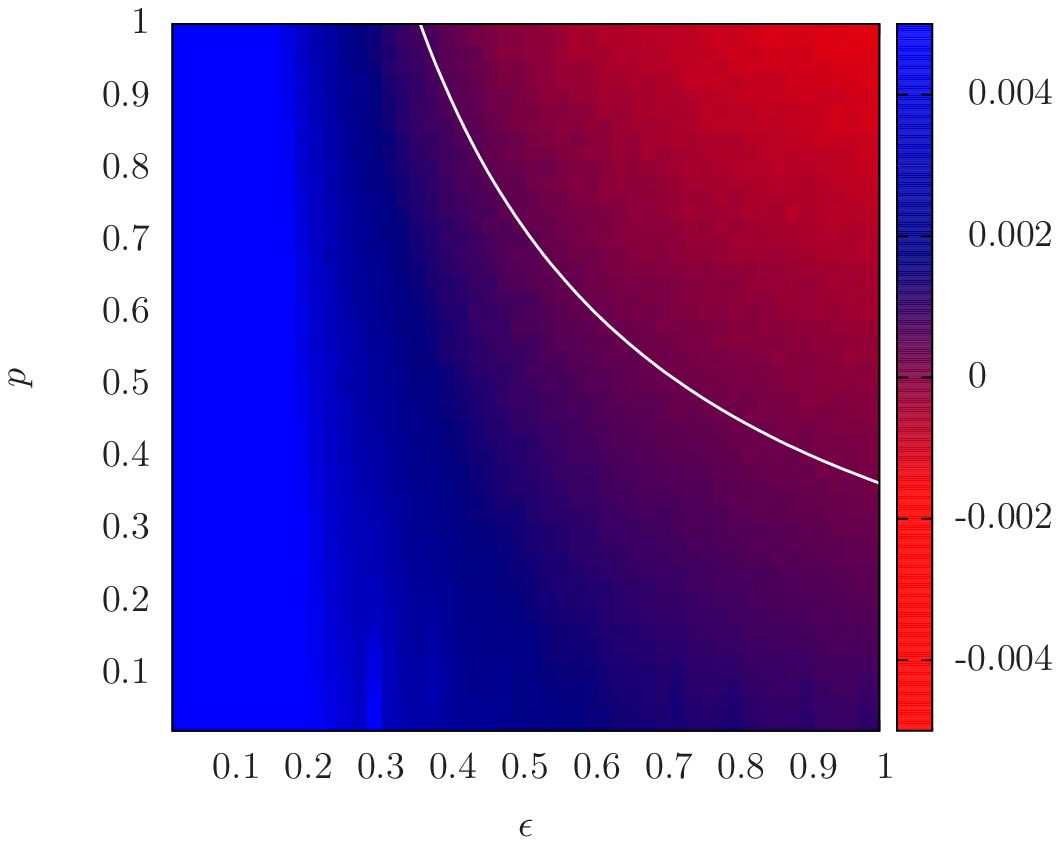,width=8cm}
\epsfig{file=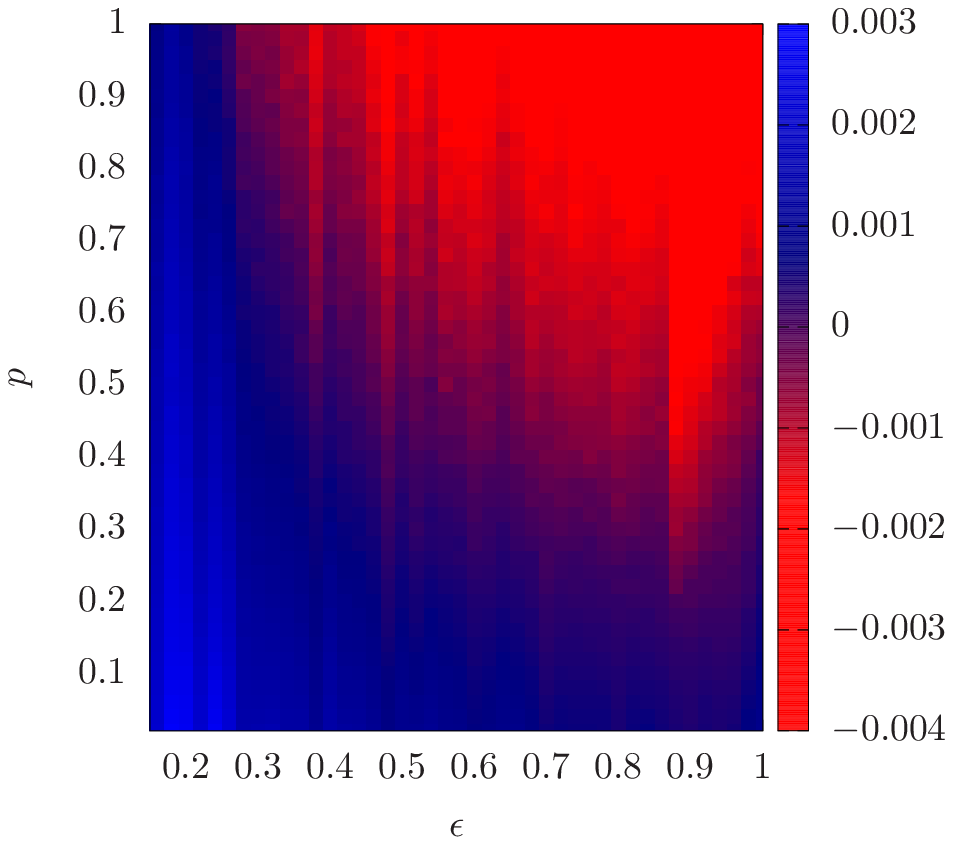,width=8cm}
\caption{(Color online) Average SLE for our SW networks as a function of both
  $\epsilon$ and $p$, using $100$ samples for each point. Top:
  Bernoulli map. Bottom: Logistic map. The white line delimits the theoretical synchronization
  region, Eq. \ref{eq:sync_bernoulli}, for a static network of Bernouilli maps and eigengap $\Delta=p/4$.}
\label{fig:sw_lyapmaps}
\end{figure}

We have also performed a thorough exploration of the $(\epsilon,p)$-parameter space. In Fig. \ref{fig:sw_lyapmaps}, we find the average value of the SLE after $100$ samples for each point, for the Bernoulli (top) and for the Logistic (bottom) maps. In both figures,
red represents negative SLE, which allow for stable synchronization,
while blue stands for positive values, which drive the system away from the synchronized state. The white line  represents the theoretical
synchronization line for networks of Bernoulli maps Eq. \eqref{eq:sync_bernoulli} for an eigenvalue gap of $\Delta=p/4$, and follows the zero average SLE accurately.

\section{Fluctuating networks}
\label{sec:fluctuating}

Let us allow our networks to fluctuate, making $G$ time-dependent with a
network switching period $T_n$: the coupling
topology will switch from the current network, $G_{\text{curr}}$, to a newly
sampled $G_{\text{next}}$ every $T_n$ time steps. We  sample from the
ensemble of all SW graphs with fixed $N$ and $p$ defined in section \ref{sec:sw}. The sampled networks are then row-normalized such that the synchronized solution exists. The Bernoulli
and Logistic map parameter values for $a$ and $r$, respectively, are
the same as in the preceeding section \ref{sec:sw}.

\begin{figure}
\epsfig{file=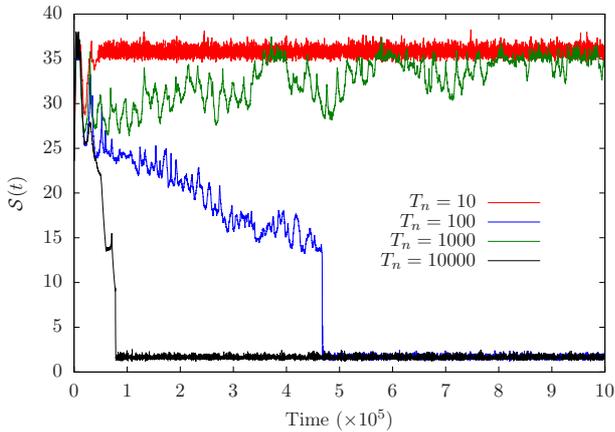,width=8cm}
\caption{(Color online) Synchronization level as a function of time for four
  histories of our dynamical system with $N=40$, $p=0.5$, $T_d=100$
  and $\epsilon=0.7$ using different fluctuation times. From upper to lower curve: $T_n=10$,
  $1000$, $100$ and $10000$, respectively. The units are Bernoulli maps.}
\label{fig:show}
\end{figure}

As a first attempt, we see in Fig. \ref{fig:show} a few time traces of the synchronization level $\cal S$ for fluctuating SW networks of Bernoulli maps with
$N=40$, $p=1/2$, $\epsilon=0.7$, and different fluctuation times: $T_n=10$, $10^2$, $10^3$ and $10^4$. For this parameters choice the average SLE is positive albeit small: $\lambda = 0.00022$. Unless otherwise stated, we will always choose $T_d=100$ for the time-delay. For $T_n=10^4$, the synchronization level decays to its minimal value  fast and full de-synchronization is irreversible. For $T_n=1000$, we observe strong fluctuations in the state deviation, but an ultimate synchronization. For $T_n=100$, when the fluctuation time-scale coincides with the delay, the system desynchronizes, although more slowly. For $T_n=10$, the system synchronizes fully quite fast.

\begin{figure}
\epsfig{file=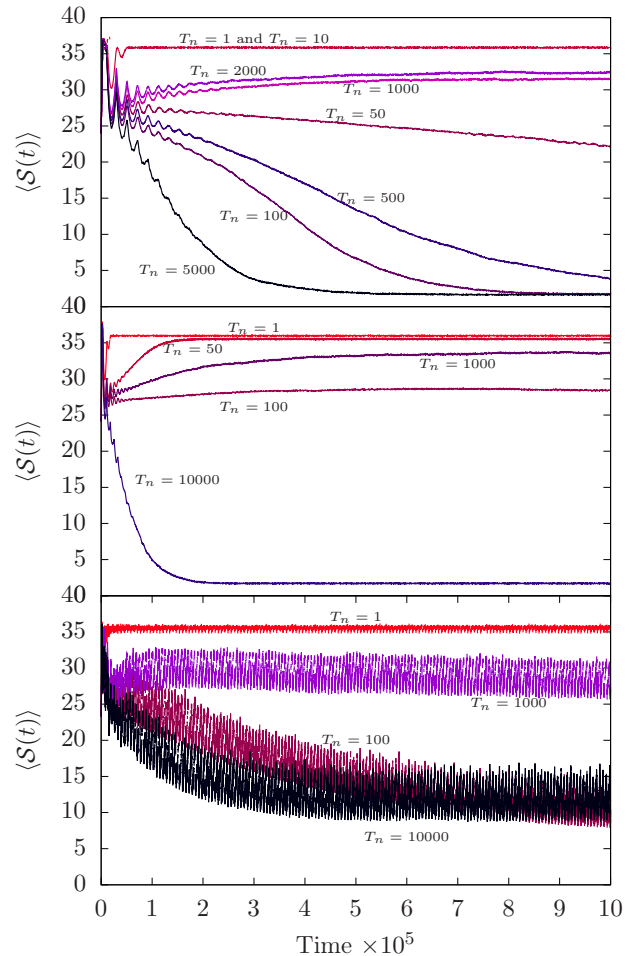,width=8cm}
\caption{(Color online) Realization average of the synchronization level, $\<{\cal
    S}(t)\>$, Eq. \eqref{eq:sync_level}, over $N_s=1000$ samples, as a
  function of time for different systems. Top: Bernoulli system,
  $N=40$, $p=0.5$ and $\epsilon=0.7$ (in average, non-synchronizing),
  for different values of $T_n$. Center: Bernoulli system with $N=40$
  sites, $p=0.8$ and $\epsilon=0.47$ (in average, synchronizing), for
  different values of $T_n$. Bottom: Logistic system, $N=40$, $p=0.5$
  and $\epsilon=0.4$ (in average, non-synchronizing). In all cases,
  $T_d=100$, and we only show time steps which are multiples of
  $100$.}
\label{fig:average}
\end{figure}

Let us average the synchronization level for a large number of realizations, $N_s=1000$. Fig. \ref{fig:average} shows the results for three different systems of coupled Bernoulli maps, showing only time steps which are multiples of $T_d=100$. Fig. \ref{fig:average} (top) shows the average synchronization for the same system as in Fig. \ref{fig:show}, in order to assess whether those results are generic. We see that for very short fluctuation time $T_n=1$ or $10$, the average synchronization level remains close to its maximum value. Hence the system synchronizes completely for all realizations independently of whether the instantaneous networks are synchronizing or not. This result for the fast fluctuation regime is quite general in all cases we considered and we discuss it on the next section. For $T_n=50$ the synchronization level decays very slowly with time, and it decays much faster for $T_n=100$. As we increase the fluctuation time, for $T_n=500$, the synchronization decay is again a bit slower, and for
$T_n=1000$ or $2000$ we can see that the system does not seem to desynchronize, but stays at a lower level of synchronization. For even slower fluctuations, $T_n=5000$, the system
desynchronizes again quite fast. We also remark the presence of
oscillatory behavior for short times.

Fig. \ref{fig:average} (center) shows the case of a Bernoulli system over SW networks with $N=40$, $p=0.8$ and $\epsilon=0.47$, a parameter choice for which the average SLE is negative. The main difference that we can observe is that for $T_n=100$ the system does not desynchronize, even though the asymptotic synchronization level is lower than for $T_n=50$ and $T_n=1000$. However, for $T_n=10000$ the system still desynchronizes. The short time oscillations that we observed in the previous case are still present, but attenuated. One may ask how is it possible that, if the average SLE is syncrhonizing, the system desynchronizes in the long run for slow network fluctuations. The reason is that, for Bernoulli systems, desynchronization is {\em irreversible}. Once a non-synchronizing network is sampled the system will start to desynchronize. If the system crosses a certain $\cal S$ threshold the subsequently sampled synchronizing networks cannot take the system back to synchronization. 

The lowest panel of Fig. \ref{fig:average} shows the average synchronization level for a Logistic system on fluctuating SW networks with $N=40$, $p=1/2$ and $\epsilon=0.4$, a parameter choice for which the average SLE is positive. The system synchronizes for fast network fluctuations $T_n=10$, and for intermediate network times $T_n=1000$, but not when network fluctuation time and time-delay coincide. Again, in the slow fluctuations regime, the synchronized state is unstable. Finally, we report much stronger oscillations for this case than for Bernouilli.

\subsection{Critical values of $\epsilon$}

\begin{figure}
\epsfig{file=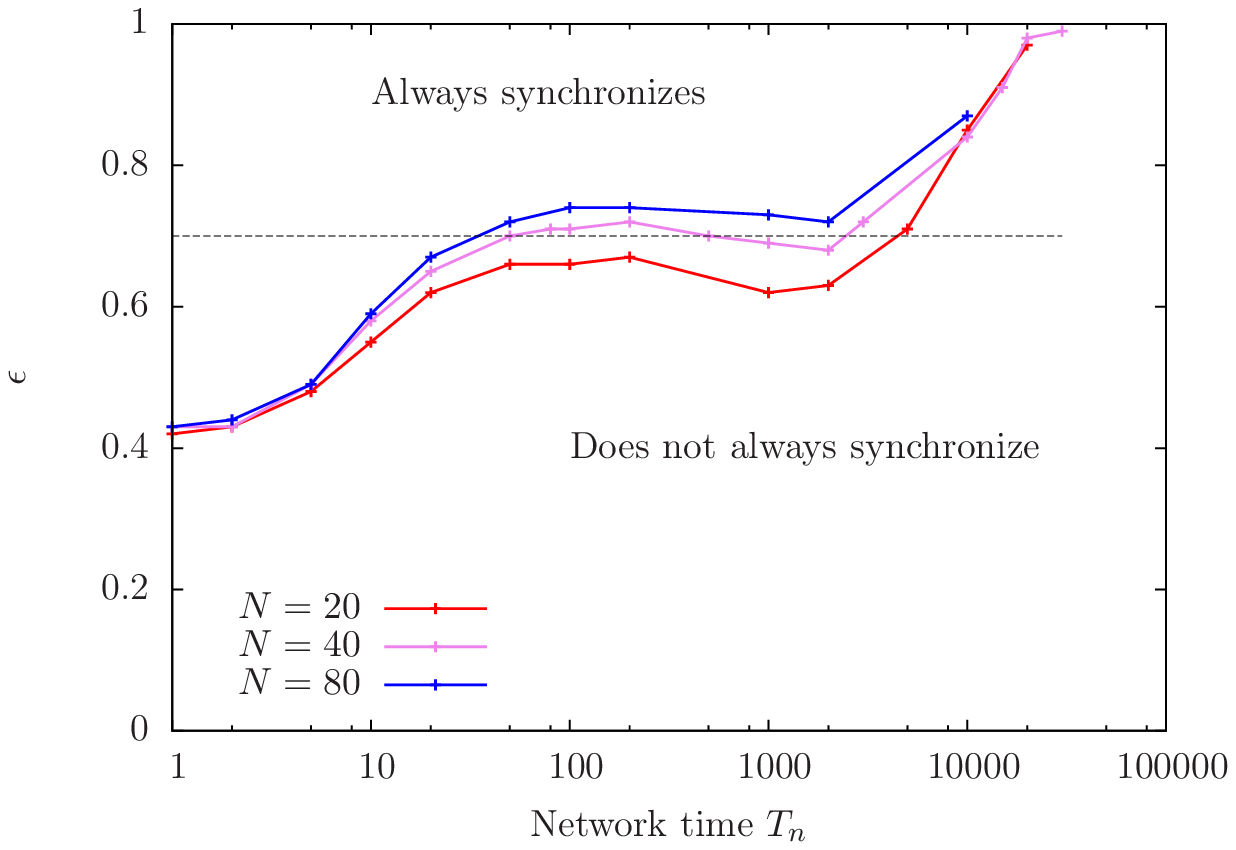,width=8cm}
\epsfig{file=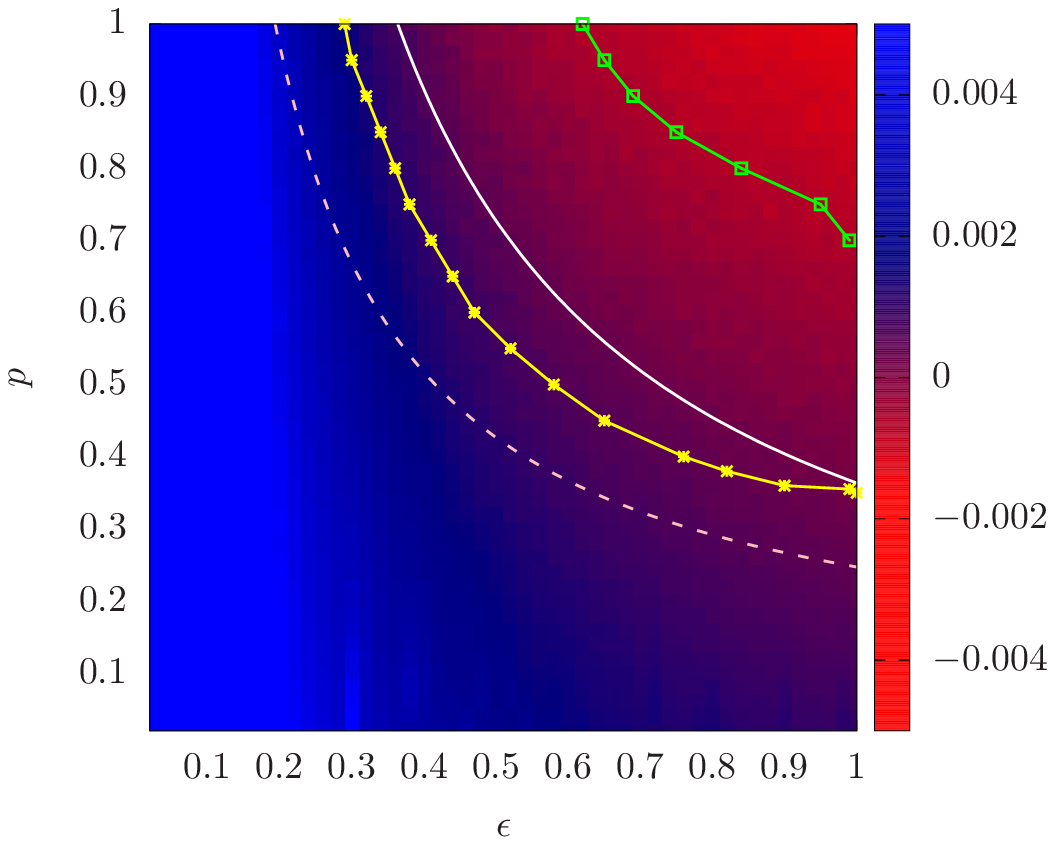,width=8cm}
\caption{(Color online) Top: Critical value $\epsilon^*$ as a function of $T_n$ for
  different values of $N=20$, $40$ and $80$ for the lower, intermediate and upper curves, respectively. 
  In all cases, $T_d=100$ and $p=1/2$, and the dynamical system is Bernoulli. The horizontal
  dashed line corresponds to the results shown in
  Fig. \ref{fig:average} (center), $\epsilon=0.7$. Bottom: same SLE
  plot as a function of $\epsilon$ and $p$ as in
  Fig. \ref{fig:sw_lyapmaps}, but adding extra lines. The yellow, full-squares line
  denotes the $\epsilon^*$ values as a function of $p$ for very fast
  fluctuations, $T_n=10$. Notice that even systems with positive
  average SLE will synchronize for those fast fluctuations. For
  comparison, the green, open-squares line denotes the $\epsilon^*$ value for very
  slow fluctuations, $T_n=10^4$. The dashed pink line denotes the
  mean-field approximation given by Eq. \eqref{eq:meanfieldresult}. The
  continuous white line corresponds to the theoretical synchronization region of a
  static network, from Eq. \ref{eq:sync_bernoulli} and is included for comparison with the non fluctuating case.}
\label{fig:epscrit}
\end{figure}

We have found out that, with all other parameters fixed, there is
always a critical value of $\epsilon$, which we call $\epsilon^*$ such
that, if $\epsilon>\epsilon^*$ the system stays synchronized almost
surely, meaning that among the $N_s=1000$ samples launched, all of
them stayed synchronized up to time $10^6$. Fig. \ref{fig:epscrit}
shows the critical $\epsilon^*$ for Bernoulli systems as a function of
the network fluctuation time, using fixed values for the other
parameters, $T_d=100$ and $p=1/2$. The three curves correspond to
different system sizes $N=20$, $40$ and $80$.

Notice that, for $N=40$ and $\epsilon=0.7$ the transition curve
presents a reentrant behavior: the critical curve is crossed three
times (see dashed horizontal line in Fig. \ref{fig:epscrit}). This is in agreement with the peculiar behavior found in the synchronization level
averages in Fig. \ref{fig:average}. 

But Fig. \ref{fig:epscrit} (top) provides more information. In all cases,
$\epsilon^*$ is minimal for small $T_n$. This implies that
synchronization is most likely to happen for fast network
fluctuations. Also, for very large $T_n$ the values of $\epsilon^*$
tend to one: synchronization becomes nearly
impossible. For network fluctuation times $T_n \sim T_d$ we have an
interesting increase in the value of $\epsilon^*$, thus implying that
when {\em both time-scales collide, synchronizability is lower}. This
phenomenon bears similarity to observations reported on neural network
models with spike-timing dependent plasticity \cite{knoblauch2012}.

The bottom panel of Fig. \ref{fig:epscrit} shows the same average
(static) SLE as a function of $\epsilon$ and $p$ for a Bernoulli
system with $N=40$ and $T_d=100$, marking also in white the zero SLE
line, i.e.: synchronization in average. The yellow, full-squares line marks the
critical $\epsilon^*$ line of almost-sure synchronization for
$T_n=10$, which is below the average synchronization line for static
networks. This means that even values of the parameters which do not
yield synchronization in average, will still synchronize almost surely
when subject to fast fluctuations. 


This enhancement of the synchronizability for rapidly fluctuating networks has been reported for a variety of systems with diffusive coupling \cite{PeruaniMorelli2010, FujiwaraGuilera2011, FrascaBoccaletti2008}. Moreover, the fast-switching 
approximation states that when the time-scale of the network fluctuations
is much larger than the typical time-scale of the oscillator dynamics, the synchronization properties
are well described by a mean-field network \cite{StilwayRobertson2006}. This is not the case here:
the pink dashed line in Fig. \ref{fig:epscrit} (bottom) delimits the mean-field synchronization region expected from the spectrum in Eq. \eqref{eq:meanfieldresult}. 
As we can see, it differs from the yellow curve for fast network fluctuations, hence
in our setting the mean-field curve does not provide a good approximation for 
the synchronization region of the fast switching regime. 
Notice that our setting is different from the classic fast switching approximation framework, which
was developed for diffusive coupling. Instead, our couplings are given by the stochastic adjacency matrix 
and not by the Laplacian. Also, the interpretation of the internal time scale $T_d$ for time-delayed systems is not so simple as for the case of non-delayed oscillators. The weak chaos regime demands $T_d\gg1$ and the necessary time to reach synchronization is given by the SLE, which scales inversely with $T_d$. Hence, for positive SLE a larger internal time-scale implies faster decay to synchronization. Nontheless, the behaviour is qualitatively similar to the fast switching approximation and the synchronizability is enhanced for $T_n \ll T_d$.

Finally, the green, open-squares line shows the $\epsilon^*$ line for $T_n=10^4$. This corresponds to the
static network regime $T_n \gg T_d$. The synchronization region is much smaller: for slow fluctuations, it is very difficult to force the system to synchronize.

\begin{figure}
\epsfig{file=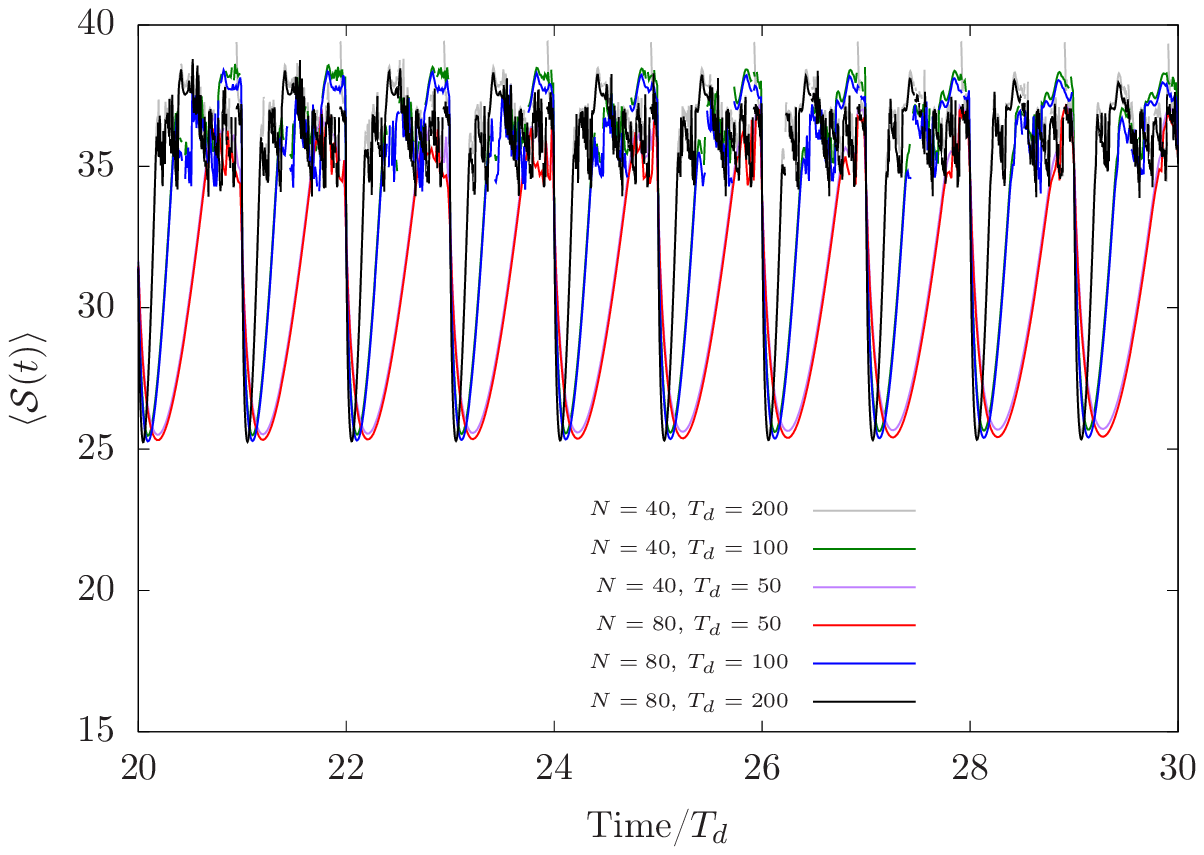,width=8cm}
\epsfig{file=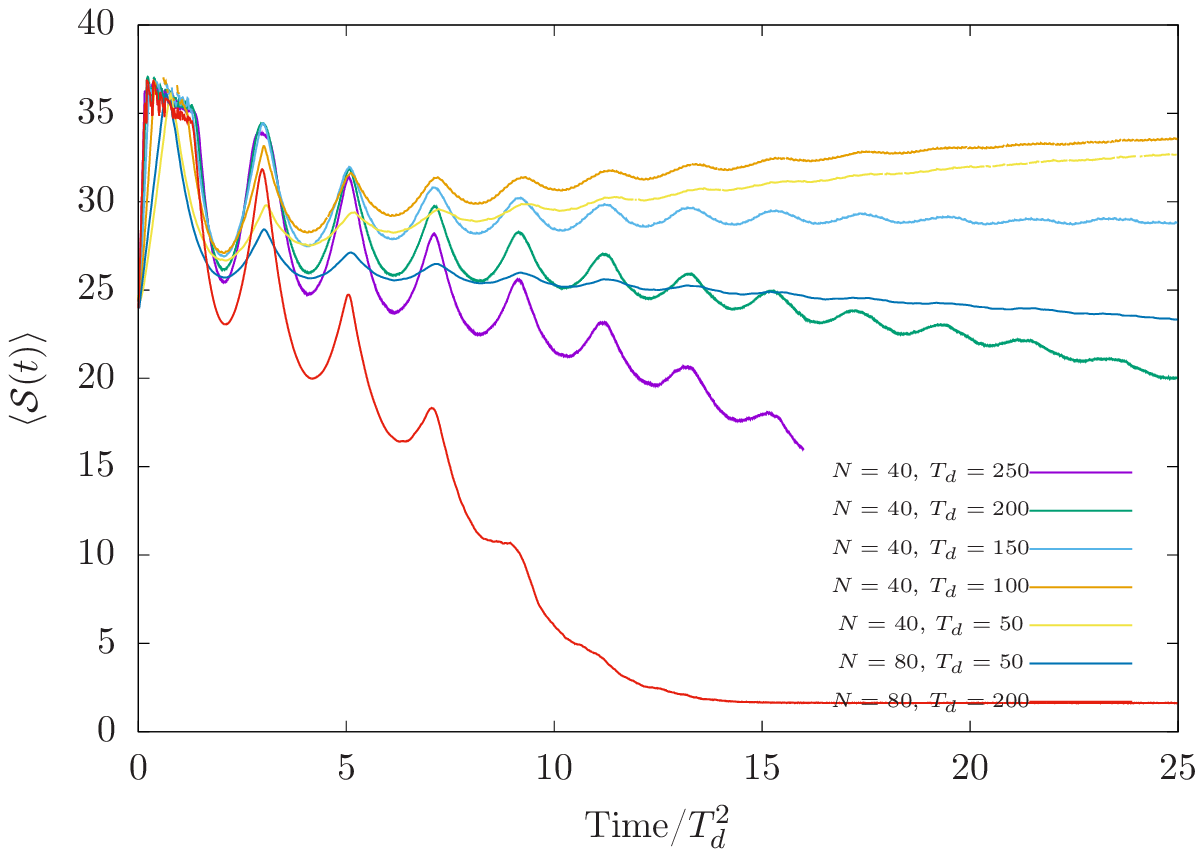,width=8cm}
\epsfig{file=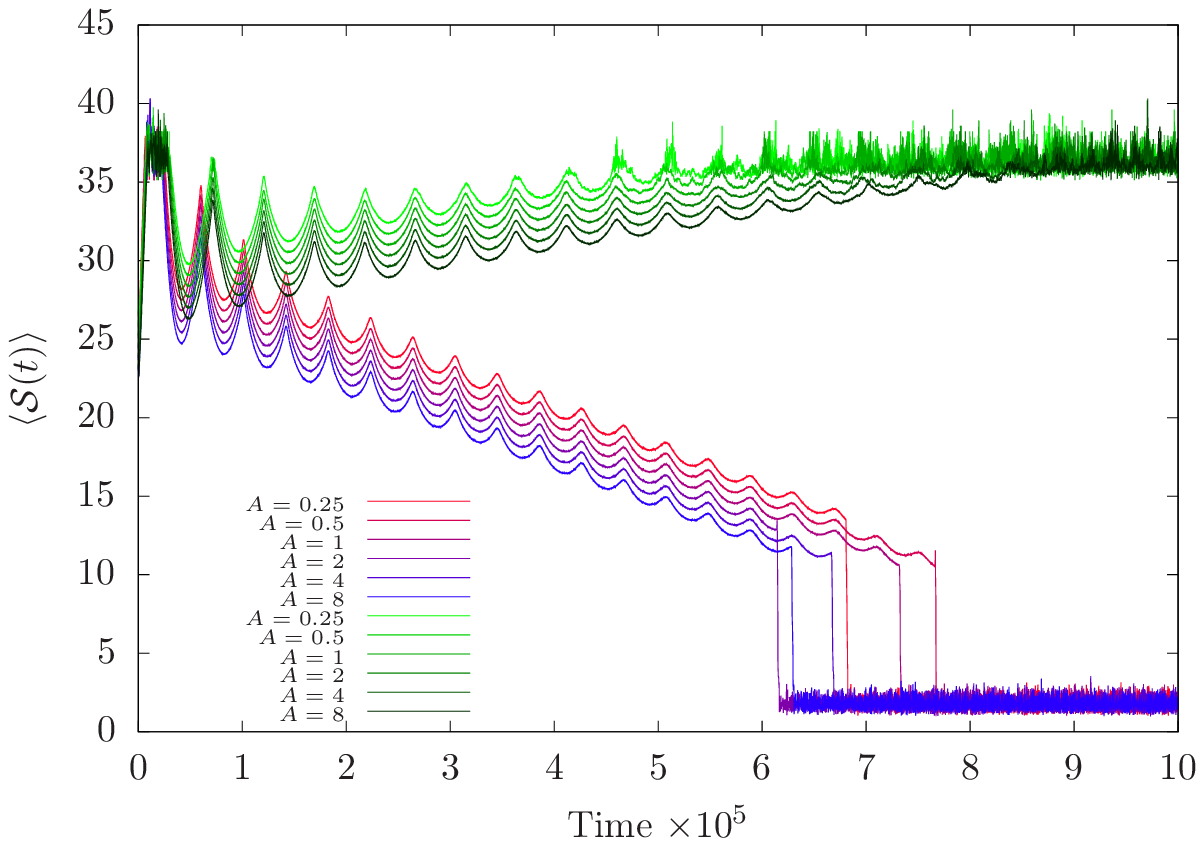,width=8cm}
\caption{(Color online) Top: Short-time evolution of the average synchronization
  level, with time in units of $T_d$, for $p=0.5$, for a Bernouilli system with $\epsilon=0.7$ and
  $T_n=1000$, using different values of $T_d$ and $N$. Center: Long
  time evolution, with the short period $T_d$ filtered out and time in
  units of $T_d^2$. Notice that, in all cases, the oscillations have
  the same frequency and phase. Bottom: Average synchronization level
  for two fixed networks with $N=20$, $p=0.7$ and $\epsilon=0.83$. The
  ascending greenish sequence corresponds to a synchronizing instance, while the
  descending bluish sequence refers to a non-synchronizing one. In both cases, we
  have averaged over random initial perturbations with amplitude
  $A\cdot 10^{-10}$. Notice that, in all cases, we obtain oscillations
  with the same frequency and phase, but different amplitudes.}
\label{fig:osc}
\end{figure}

\subsection{Synchronization oscillations}
A very salient feature of the average synchronization level curves in Fig. \ref{fig:average} is the presence of oscillations, which decay with time. The oscillations all show a periodicity related to the time-delay $T_d$, and they have the same phase for different values of the network switching time $T_n$. We illustrated this in further detail in Fig. \ref{fig:osc} (top), which shows the average synchronization level for a short time-span for all time-steps (not only multiples of 100) using different values of $N$ and $T_d$. These oscillations are independent of the network fluctuations, and they appear as well in fixed networks, as shown in Fig. \ref{fig:osc} (bottom). 

Upon this primary oscillation we have found a secondary oscillation, which periodicity scales with the square of the time-delay, $T_d^2$. In Fig. \ref{fig:osc} (center) we have removed the primary period, by showing only times multiple of $T_d$, and we represent a longer time span, with the time axis rescaled to $t/T_d^2$. Again, these fluctuations have the same frequency and phase in all cases. 

These oscillations are related to our choice of initial condition: we perturb the system at $t=0$, and this perturbation decays initially. The initial decay is a typical behavior for weakly chaotic systems \cite{DHuysKinzel2013}. After a delay time the perturbation reappears, and this reappearance of a delay echo can be related to the observed periodicity equal to the delay time. For Bernoulli maps the evolution can even be calculated explicitly, we included the analytic calculations in appendix \ref{ap:anal}.

\begin{figure}
\epsfig{file=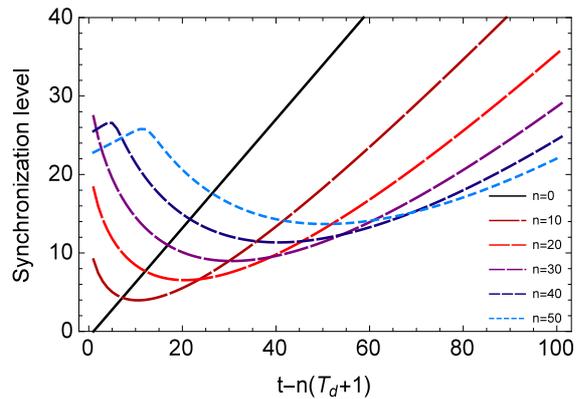,width=7.5cm}
\caption{(Color online) Delay echoes of a point-like perturbation applied at $t=0$ along the transverse direction $v_2$, in a network of delay-coupled Bernoulli maps. We show the synchronization level $-\ln|v_2(t)|$ immediately after applying the perturbation, and the tenth, twentieth, thirtieth, fortieth and fiftieth delay echo. We normalized with respect to the initial amplitude. Parameters are $T_d=100$, $a=3/2$, $\epsilon=2/3$, $\gamma_2=2/5$}
\label{fig:delayechos}
\end{figure}

This delay echo is transformed each time-delay interval and its shape gradually changes from an exponential decay to decaying oscillatory motion. In Fig. \ref{fig:delayechos} we show the analytically calculated evolution of a point-like perturbation along a specific direction in a fixed network of Bernoulli elements. While the initial perturbation (blue line) decays exponentially, the consecutive delay echoes are broader, and reach their maximal amplitude at a later point in time within the delay interval. However, the exact mathematical origin of the observed secondary oscillations remains to be explained.

For general initial conditions in the vicinity of the synchronization manifold, one observes as well one or more frequency components related to the time-delay, and we conjecture that they are characteristic for delay systems. We expect however that the phase of the oscillations (and thus the fact that they do not average out over multiple instances) is a result of our choice of initial condition.

\begin{figure}[t]
\epsfig{file=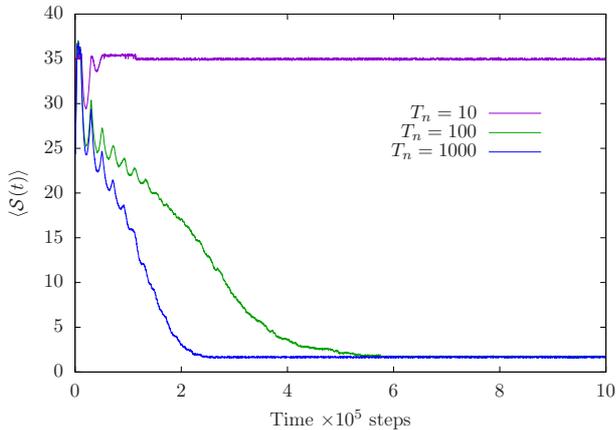,width=8cm}
\caption{(Color online) Average synchronization for Bernoulli systems with
  $\epsilon=0.7$ and $T_d=100$ on SW networks with $N=40$, $p=1/2$ and
  the additional constraint $\Delta < 1/10$ for {\em every}  sampled network. 
  The probability to synchronize for this spectral gap values is negligible. Depicted curves 
  correspond, from upper to lower, to $T_n=10$, $100$ and $1000$.}
\label{fig:lowgap}
\end{figure}

\subsection{Low-gap networks}

In order to explain the synchronization enhancement via network
fluctuations we might conjecture that large gap networks pull the
system towards the synchronization manifold, while low gap ones push
it away. We have checked numerically that conjecture and found it to
be inaccurate. In Fig. \ref{fig:lowgap} we consider SW networks with
$N=40$ and $p=1/2$, on which we set up a Bernoulli interacting system
with $T_d=100$ and $\epsilon=0.7$. The difference is that the network
fluctuations are only allowed to explore the ensemble of {\em low gap}
graphs. Specifically, we reject all SW networks whose gap is larger
than $\Delta_*=0.1$. Approximately, the probability of rejection with
these parameters is $1/2$. Nonetheless, the probability of one of
these networks to synchronize is negligible ($<10^{-6}$ in our
numerical experiments). Fig. \ref{fig:lowgap} shows the average
synchronization for $100$ realizations, using $T_n=10$, $100$ and
$1000$, and found that, for fast enough fluctuations, the system {\em
  synchronizes}. Again, the same oscillations can be seen.
  This last result is reminiscent of the {\em Parrondo games}
\cite{Harmer1999}, where the alternation of losing strategies can give
rise to a winning one. A random alternation of non-synchronizing
networks can result strongly synchronizing.


\section{Conclusions and Further work}
\label{sec:conclusions}

The possibility to enhance the stability of a system through fast
oscillations or fluctuations is a topic of long tradition, e.g. the
Kapitza pendulum \cite{LandauMechanics}. In this work we have explored the effect of topology fluctuations on the synchronizability of small-world networks of time-delayed coupled chaotic maps. We have first studied synchronizability of static networks sampled from the Newman-Watts small world network ensemble with $N$ nodes and a fraction $p$ of shortcuts. The spectral gap was found to be approximately given by $p/4$  independently of $N$ and it showed a clear relationship with the synchronization Lyapunov exponent for networks of Bernouilli and Logistic maps. The Bernouilli map case followed closely the theoretical prediction, while the mapping was more nonlinear for the Logistc maps.

We then studied how the synchronization properties are affected by a time varying coupling topology. We found the stability of the synchronized state to be strongly affected by the interplay between the time-scale of the delayed interactions, $T_d$, and that of the network fluctuations, $T_n$. For the fast-switching regime, $T_n \ll T_d$, we obtain a strong enhancement of the synchronizability of the network. Even when we restrict our topology fluctuations to only explore those networks which would not be able to synchronize under static conditions, we can obtain almost-sure synchronization under fast enough fluctuations. This result is in qualitative agreement with the fast switching approximation \cite{StilwayRobertson2006}. For $T_n \sim T_d$ we observe a severe reduction in synchronizability, which is recovered as we increase $T_n>T_d$. Nonetheless, for $T_n\gg T_d$ the system will nearly always desynchronize. Moreover, we observe oscillations in the synchronization level, when the network is close to the synchronized state. These oscillations have a periodicity related to $T_d$ and are typical of the weak chaos regime. For our choice of perturbation, these oscillations can be analytically recovered in a network of Bernoulli maps. We also report a secondary oscillation of periodicity scaling with $T_d^2$. 

We have restricted ourselves to the case of small-world networks,
because they are more amenable to a mean-field approach, but it is
relevant to ask whether these results also apply to other network
ensembles, such as purely random Erd\H{o}s-R\'enyi graphs or scale-free networks; as well as to other dynamical systems beyond Bernoulli or Logistic.

\begin{acknowledgments}
We would like to acknowledge W. Kinzel and A. Dea\~no. This
work was partly supported by the Spanish Government through grant
FIS-2012-38866-C05-1 (J.R.-L.) and the Alexander von Humboldt
Foundation within the Renewed research stay program (E.K.).
\end{acknowledgments}


\appendix
\section{Mean-field spectrum and eigengap of the SW networks}
\label{ap:MF}

Our ensemble is composed of networks which consist of a directed ring
of $N$ nodes to which we add $Np$ directed shortcuts. Here we compute
the spectrum of the networks of this ensemble within the mean-field
approximation of \cite{GrabowJurgenPRE2015}, in order to characterize
the eigengap. The strategy is as follows: we obtain the ensemble
average of each adjacency matrix entry, $\<G_{ij}\>$, and study the
spectrum of the resulting matrix, which will be a {\em circulant
  matrix},

\begin{equation}
G^{\textsc{mf}}=
\begin{bmatrix}
  c_0 & c_{N-1} & \dots & c_2 & c_1 \\
  c_1 & c_0 & \dots & c_3 & c_2 \\
  \vdots & \vdots & \ddots & \vdots & \vdots \\
  c_{N-1} & c_{N-2} & \dots & c_1 & c_0  
\end{bmatrix},
\label{eq:circulant}
\end{equation}
whose eigenvalues can be computed analytically \cite{GolubVanLoan1996}:

\begin{equation}
  \gamma_m^{\textsc{mf}} = \sum_k c_k \exp\left[ \frac{-2\pi i m k}{N}\right],
  \label{eq:circulant_spectrum}
\end{equation}
where $m=0,\dots,N-1$. Before the shortcuts are introduced, $c_1=1$
and all other entries $c_i=0$, $i\neq 1$. When the shortcuts are
introduced, the subdiagonal elements become $G_{i,i+1}=1/(1+n_s)$,
where $n_s$ is the number of shortcuts reaching element $i$. Thus,

\begin{equation}
  c_1 = \<G_{i,i+1}\> = \< \frac{1}{1+n_s} \>.
  \label{eq:c1average}
\end{equation}
The random variable $n_s$ follows a binomial distribution: each site
can be reached by $\approx N$ possible shortcuts, each of them with
probability $\approx p/N$. Thus, its probability distribution is given
by

\begin{equation}
  P(n_s) \approx {N \choose n_s} \({p\over N}\)^{n_s} \( 1-{p\over
    N}\)^{N-n_s},
  \label{eq:binomial}
\end{equation}
which, in the limit where $p/N \ll 1$ approaches the Poisson
distribution

\begin{equation}
  P(n_s) \approx e^{-p}\; {p^{n_s}\over n_s!}.
  \label{eq:poisson}
\end{equation}
For this distribution it is possible to obtain the desired expected
value:

\begin{align}
  c_1 &= \< {1\over 1+n_s} \> \approx \sum_{n_s=0}^\infty {1\over 1+n_s}\;
  e^{-p}\; {p^{n_s}\over n_s!} \nonumber \\
  &={e^{-p}\over p} \sum_{x=1}^\infty {p^x\over x!} \nonumber \\
  &= {1-e^{-p}\over p},
  \label{eq:expvalue}
\end{align}
where we choose $x=n_s+1$.  The rest of the entries of the circulant
matrix are all equal, $c_i=\tilde c$ for $i\neq 1$, and can be found
by normalization:

\begin{equation}
  \tilde c \approx {1\over N-1}\; \(1- {1-e^{-p}\over p}\). 
  \label{eq:cifinal}
\end{equation}

Thus, applying \eqref{eq:circulant_spectrum} we have

\begin{equation}
  \gamma_m^{\textsc{mf}} = 
  \tilde c + c_1 \e^{\frac{2\pi i m}{N}} + 
  \tilde c \sum_{k=2}^{N-1} \e^{\frac{2\pi i k m}{N}}.
\end{equation}
If $m=0$, we obtain $\gamma_0=1$. The last term can be evaluated as a
geometric sum or, alternatively, we can realize that if $k$ was
extended from $0$ to $N-1$, it would yield zero. In both cases, we
obtain 

\begin{equation}
  \gamma_m^{\textsc{mf}} = (c_1-\tilde c) \e^{-\frac{2\pi i m}{N}}.
\end{equation}

Thus, the modulus of all eigenvalues for $m>0$ is equal:

\begin{equation}
  |\gamma_m^{\textsc{mf}}| \approx c_1 - \tilde c =
  {1-e^{-p}\over p} - {1\over N-1} \( 1- {1-e^{-p}\over p} \).
\end{equation}

Neglecting corrections of order $N^{-1}$, the eigenvalue gap is:

\begin{equation}
  \Delta = 1 - \max\{|\gamma_{m\neq 0}^{\textsc{mf}}|\} \approx
         1- {1-e^{-p}\over p}.
\end{equation}

\section{Analytic explanation of the synchronization oscillations}
\label{ap:anal}
We show in this appendix how synchronization oscillations arise generically, as the network decays towards or drifts away from the synchronization manifold. We consider a network of Bernoulli maps and the network structure to be fixed. 

In this case, as the initial perturbation $\bf{\xi}(t)$ is small, we linearize around the synchronization manifold $u_i(t)=\mu(t)$. The evolution of the network is then given by
\begin{multline}
\xi_i(t+1)=(1-\epsilon)f'(\mu(t)) \xi_i(t)+ \\  \epsilon\sum_j G_{ij}f'(\mu(t-T_d)) \xi_i(t-T_d)\label{eq:pert0}\,.
\end{multline}

After decomposition along the eigenvectors $\{v_k\}$ of $G$, we can rewrite Eq. \eqref{eq:pert0} as
\begin{multline}
v_k(t+1)=(1-\epsilon) f'(\mu(t)) v_k(t)+ \\  \epsilon \gamma_k f'(\mu(t-T_d))  v_k(t-T_d)\label{eq:pert1}\,,
\end{multline}
where $\gamma_k$ denotes the eigenvalue of $G$ along the eigenvector $v_k$. For Bernoulli maps, the derivative along the chaotic trajectory is constant, $f'(u(t))=a$, we can simplify Eq. \eqref{eq:pert1} as
\begin{multline}
v_k(t+1)=(1-\epsilon) a v_k(t)+  \epsilon \gamma_k a  v_k(t-T_d)\label{eq:pert1B}\,.
\end{multline}

The exponential decay of a perturbation is slowest (or the growth is fastest) along the direction with the smallest eigenvalue gap $1-|\gamma_2|$, we thus only consider the direction $v_2(t)$. To find the evolution along a direction $v_k$, one can simply replace $\gamma_2$ by $\gamma_k$ in the calculations. In the simulations we applied a perturbation $\xi(t)$ only at $t=0$, with a randomized magnitude over the network nodes. For simplicity, we will assume this magnitude along the direction $v_2(0)=1$, while $v_2(t<0)=0$. We can then solve Eq. \eqref{eq:pert1B} directly, and we find for the first delay interval, $0 \leq t < T_d$,
\begin{eqnarray}
v_2(t+1) & = & (1-\epsilon)a v_2(t)\,,
\end{eqnarray}
which is solved by 
\begin{eqnarray}
v_2(t) & = & (\left(1-\epsilon)a\right)^t\mbox{ for } 0< t \leq T_d\label{eq:interval1}\,.
\end{eqnarray}
The perturbation initially evolves with a rate given by $\ln |(1-\epsilon)a|$, which corresponds to the instantaneous Lyapunov exponent \cite{Heiligenthal2011, DHuysKinzel2013}. We only consider networks in the weakly chaotic regime, meaning that the instantaneous Lyapunov exponent is negative and that the perturbation initially decays. 

Using Eq. \eqref{eq:interval1} as initial function for the next delay interval $T_d \leq t < 2T_d+1$, this leads to an equation of motion
\begin{eqnarray}
v_2(t+1) & = & (1-\epsilon)a v_2(t)+\epsilon a \gamma_2 (\left(1-\epsilon)a\right)^{t}\,.
\end{eqnarray}
Imposing continuity, $v_2(T_d)=\left((1-\epsilon)a\right)^{T_d}$, this difference equation is solved by
\begin{equation}
v_2(t) = \left((1-\epsilon)a\right)^t + (t-T_d)\epsilon a \gamma_2 (\left(1-\epsilon)a\right)^{t-T_d-1}\,,
\end{equation}
for $T_d<t\leq 2T_d+1$. Hence, the initial perturbation reappears after a time $T_d+1$, but the delay echo is broadened. In general, we find for $n(T_d+1)\leq t<(n+1)(T_d+1)$
\begin{eqnarray}
v_2(t) & = & \displaystyle\sum^n_{k=0} \frac{1}{k!}\left((1-\epsilon)a\right)^{t-kT_d-k}(a\epsilon\gamma_2)^k\displaystyle\prod_{l=0}^{k-1}(t-kT_d-l)\,.\nonumber\\
\end{eqnarray}

We find additional delay echoes appearing at multiples of $T_d+1$, each one broader than the previous. After several cycles the resulting motion indeed resembles an oscillation, with a periodicity approximated as $$T\approx T_d+1/2-1/\ln|a(1-\epsilon)|,$$ in the limit of large delay $T_d\rightarrow \infty$. This is illustrated in Fig. \ref{fig:delayechos}. We remark here that in general, $\gamma_2$ is a complex number, which could lead to additional oscillations with a periodicity related to the delay time and the phase of $\gamma_2$.


\section*{References}

\begin{thebibliography}{40}
\expandafter\ifx\csname natexlab\endcsname\relax\def\natexlab#1{#1}\fi
\expandafter\ifx\csname bibnamefont\endcsname\relax
  \def\bibnamefont#1{#1}\fi
\expandafter\ifx\csname bibfnamefont\endcsname\relax
  \def\bibfnamefont#1{#1}\fi
\expandafter\ifx\csname citenamefont\endcsname\relax
  \def\citenamefont#1{#1}\fi
\expandafter\ifx\csname url\endcsname\relax
  \def\url#1{\texttt{#1}}\fi
\expandafter\ifx\csname urlprefix\endcsname\relax\def\urlprefix{URL }\fi
\providecommand{\bibinfo}[2]{#2}
\providecommand{\eprint}[2][]{\url{#2}}

\bibitem[{\citenamefont{Boccaletti et~al.}(2002)\citenamefont{Boccaletti,
  Kurths, Osipov, Valladares, and Zhou}}]{Boccaletti2002}
\bibinfo{author}{\bibfnamefont{S.}~\bibnamefont{Boccaletti}},
  \bibinfo{author}{\bibfnamefont{J.}~\bibnamefont{Kurths}},
  \bibinfo{author}{\bibfnamefont{G.}~\bibnamefont{Osipov}},
  \bibinfo{author}{\bibfnamefont{D.}~\bibnamefont{Valladares}},
  \bibnamefont{and} \bibinfo{author}{\bibfnamefont{C.}~\bibnamefont{Zhou}},
  \bibinfo{journal}{Physics Reports} \textbf{\bibinfo{volume}{366}},
  \bibinfo{pages}{1 } (\bibinfo{year}{2002}).

\bibitem[{\citenamefont{Pecora and Carroll}(1998)}]{Pecora1998}
\bibinfo{author}{\bibfnamefont{L.~M.} \bibnamefont{Pecora}} \bibnamefont{and}
  \bibinfo{author}{\bibfnamefont{T.~L.} \bibnamefont{Carroll}},
  \bibinfo{journal}{Phys. Rev. Lett.} \textbf{\bibinfo{volume}{80}},
  \bibinfo{pages}{2109} (\bibinfo{year}{1998}).

\bibitem[{\citenamefont{Flunkert et~al.}(2010)\citenamefont{Flunkert, Yanchuk,
  Dahms, and Sch\"oll}}]{Scholl2010}
\bibinfo{author}{\bibfnamefont{V.}~\bibnamefont{Flunkert}},
  \bibinfo{author}{\bibfnamefont{S.}~\bibnamefont{Yanchuk}},
  \bibinfo{author}{\bibfnamefont{T.}~\bibnamefont{Dahms}}, \bibnamefont{and}
  \bibinfo{author}{\bibfnamefont{E.}~\bibnamefont{Sch\"oll}},
  \bibinfo{journal}{Phys. Rev. Lett.} \textbf{\bibinfo{volume}{105}},
  \bibinfo{pages}{254101} (\bibinfo{year}{2010}).

\bibitem[{\citenamefont{Atay et~al.}(2004)\citenamefont{Atay, Jost, and
  Wende}}]{Atay2004}
\bibinfo{author}{\bibfnamefont{F.~M.} \bibnamefont{Atay}},
  \bibinfo{author}{\bibfnamefont{J.}~\bibnamefont{Jost}}, \bibnamefont{and}
  \bibinfo{author}{\bibfnamefont{A.}~\bibnamefont{Wende}},
  \bibinfo{journal}{Phys. Rev. Lett.} \textbf{\bibinfo{volume}{92}},
  \bibinfo{pages}{144101} (\bibinfo{year}{2004}).

\bibitem[{\citenamefont{Heil et~al.}(2001)\citenamefont{Heil, Fischer,
  Els{\"{a}}{\ss}er, Mulet, and Mirasso}}]{Heiligenthal2001}
\bibinfo{author}{\bibfnamefont{T.}~\bibnamefont{Heil}},
  \bibinfo{author}{\bibfnamefont{I.}~\bibnamefont{Fischer}},
  \bibinfo{author}{\bibfnamefont{W.}~\bibnamefont{Els{\"{a}}{\ss}er}},
  \bibinfo{author}{\bibfnamefont{J.}~\bibnamefont{Mulet}}, \bibnamefont{and}
  \bibinfo{author}{\bibfnamefont{C.}~\bibnamefont{Mirasso}},
  \bibinfo{journal}{Phys. Rev. Lett.} \textbf{\bibinfo{volume}{86}},
  \bibinfo{pages}{795} (\bibinfo{year}{2001}).

\bibitem[{\citenamefont{Fischer et~al.}(2006)\citenamefont{Fischer, Vicente,
  Buldu, Peil, Mirasso, Torrent, and Garcia-Ojalvo}}]{Fischer06}
\bibinfo{author}{\bibfnamefont{I.}~\bibnamefont{Fischer}},
  \bibinfo{author}{\bibfnamefont{R.}~\bibnamefont{Vicente}},
  \bibinfo{author}{\bibfnamefont{J.}~\bibnamefont{Buldu}},
  \bibinfo{author}{\bibfnamefont{M.}~\bibnamefont{Peil}},
  \bibinfo{author}{\bibfnamefont{C.}~\bibnamefont{Mirasso}},
  \bibinfo{author}{\bibfnamefont{M.}~\bibnamefont{Torrent}}, \bibnamefont{and}
  \bibinfo{author}{\bibfnamefont{J.}~\bibnamefont{Garcia-Ojalvo}},
  \bibinfo{journal}{Phys. Rev. Lett.} \textbf{\bibinfo{volume}{97}},
  \bibinfo{pages}{123902} (\bibinfo{year}{2006}).

\bibitem[{\citenamefont{Heiligenthal et~al.}(2011)\citenamefont{Heiligenthal,
  Dahms, Yanchuk, J\"ungling, Flunkert, Kanter, Sch\"oll, and
  Kinzel}}]{Heiligenthal2011}
\bibinfo{author}{\bibfnamefont{S.}~\bibnamefont{Heiligenthal}},
  \bibinfo{author}{\bibfnamefont{T.}~\bibnamefont{Dahms}},
  \bibinfo{author}{\bibfnamefont{S.}~\bibnamefont{Yanchuk}},
  \bibinfo{author}{\bibfnamefont{T.}~\bibnamefont{J\"ungling}},
  \bibinfo{author}{\bibfnamefont{V.}~\bibnamefont{Flunkert}},
  \bibinfo{author}{\bibfnamefont{I.}~\bibnamefont{Kanter}},
  \bibinfo{author}{\bibfnamefont{E.}~\bibnamefont{Sch\"oll}}, \bibnamefont{and}
  \bibinfo{author}{\bibfnamefont{W.}~\bibnamefont{Kinzel}},
  \bibinfo{journal}{Phys. Rev. Lett.} \textbf{\bibinfo{volume}{107}},
  \bibinfo{pages}{234102} (\bibinfo{year}{2011}).

\bibitem[{\citenamefont{Nixon et~al.}(2012)\citenamefont{Nixon, Fridman, Ronen,
  Friesem, Davidson, and Kanter}}]{Nixon2012}
\bibinfo{author}{\bibfnamefont{M.}~\bibnamefont{Nixon}},
  \bibinfo{author}{\bibfnamefont{M.}~\bibnamefont{Fridman}},
  \bibinfo{author}{\bibfnamefont{E.}~\bibnamefont{Ronen}},
  \bibinfo{author}{\bibfnamefont{A.~A.} \bibnamefont{Friesem}},
  \bibinfo{author}{\bibfnamefont{N.}~\bibnamefont{Davidson}}, \bibnamefont{and}
  \bibinfo{author}{\bibfnamefont{I.}~\bibnamefont{Kanter}},
  \bibinfo{journal}{Phys. Rev. Lett.} \textbf{\bibinfo{volume}{108}},
  \bibinfo{pages}{214101} (\bibinfo{year}{2012}).

\bibitem[{\citenamefont{Argyris et~al.}(2005)\citenamefont{Argyris, Syvridis,
  Larger, Annovazzi-Lodi, Colet, Fischer, García-Ojalvo, Mirasso, Pesquera,
  and Shore}}]{Shore2005}
\bibinfo{author}{\bibfnamefont{A.}~\bibnamefont{Argyris}},
  \bibinfo{author}{\bibfnamefont{D.}~\bibnamefont{Syvridis}},
  \bibinfo{author}{\bibfnamefont{L.}~\bibnamefont{Larger}},
  \bibinfo{author}{\bibfnamefont{V.}~\bibnamefont{Annovazzi-Lodi}},
  \bibinfo{author}{\bibfnamefont{P.}~\bibnamefont{Colet}},
  \bibinfo{author}{\bibfnamefont{I.}~\bibnamefont{Fischer}},
  \bibinfo{author}{\bibfnamefont{J.}~\bibnamefont{García-Ojalvo}},
  \bibinfo{author}{\bibfnamefont{C.~R.} \bibnamefont{Mirasso}},
  \bibinfo{author}{\bibfnamefont{L.}~\bibnamefont{Pesquera}}, \bibnamefont{and}
  \bibinfo{author}{\bibfnamefont{K.~A.} \bibnamefont{Shore}},
  \bibinfo{journal}{Nature} \textbf{\bibinfo{volume}{438}},
  \bibinfo{pages}{343} (\bibinfo{year}{2005}).

\bibitem[{\citenamefont{Kanter et~al.}(2008)\citenamefont{Kanter, Kopelowitz,
  and Kinzel}}]{KanterKinzel2008}
\bibinfo{author}{\bibfnamefont{I.}~\bibnamefont{Kanter}},
  \bibinfo{author}{\bibfnamefont{E.}~\bibnamefont{Kopelowitz}},
  \bibnamefont{and} \bibinfo{author}{\bibfnamefont{W.}~\bibnamefont{Kinzel}},
  \bibinfo{journal}{Phys. Rev. Lett.} \textbf{\bibinfo{volume}{101}},
  \bibinfo{pages}{084102} (\bibinfo{year}{2008}).

\bibitem[{\citenamefont{Buzsaki}(2006)}]{Buzsaki2009}
\bibinfo{author}{\bibfnamefont{G.}~\bibnamefont{Buzsaki}},
  \emph{\bibinfo{title}{Rhythms of the brain}} (\bibinfo{publisher}{Oxford
  University Press}, \bibinfo{year}{2006}).

\bibitem[{\citenamefont{Kanter et~al.}(2011{\natexlab{a}})\citenamefont{Kanter,
  Kopelowitz, Vardi, Zigzag, Kinzel, Abeles, and Cohen}}]{KanterCohen2011}
\bibinfo{author}{\bibfnamefont{I.}~\bibnamefont{Kanter}},
  \bibinfo{author}{\bibfnamefont{E.}~\bibnamefont{Kopelowitz}},
  \bibinfo{author}{\bibfnamefont{R.}~\bibnamefont{Vardi}},
  \bibinfo{author}{\bibfnamefont{M.}~\bibnamefont{Zigzag}},
  \bibinfo{author}{\bibfnamefont{W.}~\bibnamefont{Kinzel}},
  \bibinfo{author}{\bibfnamefont{M.}~\bibnamefont{Abeles}}, \bibnamefont{and}
  \bibinfo{author}{\bibfnamefont{D.}~\bibnamefont{Cohen}},
  \bibinfo{journal}{EPL (Europhysics Letters)} \textbf{\bibinfo{volume}{93}},
  \bibinfo{pages}{66001} (\bibinfo{year}{2011}{\natexlab{a}}).

\bibitem[{\citenamefont{Kanter et~al.}(2011{\natexlab{b}})\citenamefont{Kanter,
  Zigzag, Englert, Geissler, and Kinzel}}]{KanterKinzel2011}
\bibinfo{author}{\bibfnamefont{I.}~\bibnamefont{Kanter}},
  \bibinfo{author}{\bibfnamefont{M.}~\bibnamefont{Zigzag}},
  \bibinfo{author}{\bibfnamefont{A.}~\bibnamefont{Englert}},
  \bibinfo{author}{\bibfnamefont{F.}~\bibnamefont{Geissler}}, \bibnamefont{and}
  \bibinfo{author}{\bibfnamefont{W.}~\bibnamefont{Kinzel}},
  \bibinfo{journal}{EPL (Europhysics Letters)} \textbf{\bibinfo{volume}{93}},
  \bibinfo{pages}{60003} (\bibinfo{year}{2011}{\natexlab{b}}).

\bibitem[{\citenamefont{Martin et~al.}(2016)\citenamefont{Martin, D'Huys,
  Lauerbach, Korutcheva, and Kinzel}}]{self_citation}
\bibinfo{author}{\bibfnamefont{M.~J.} \bibnamefont{Martin}},
  \bibinfo{author}{\bibfnamefont{O.}~\bibnamefont{D'Huys}},
  \bibinfo{author}{\bibfnamefont{L.}~\bibnamefont{Lauerbach}},
  \bibinfo{author}{\bibfnamefont{E.}~\bibnamefont{Korutcheva}},
  \bibnamefont{and} \bibinfo{author}{\bibfnamefont{W.}~\bibnamefont{Kinzel}},
  \bibinfo{journal}{Phys. Rev. E} \textbf{\bibinfo{volume}{93}},
  \bibinfo{pages}{022206} (\bibinfo{year}{2016}).

\bibitem[{\citenamefont{Feng et~al.}(2006)\citenamefont{Feng, Jirsa, and
  Ding}}]{FengMingzhouC2006}
\bibinfo{author}{\bibfnamefont{J.}~\bibnamefont{Feng}},
  \bibinfo{author}{\bibfnamefont{V.~K.} \bibnamefont{Jirsa}}, \bibnamefont{and}
  \bibinfo{author}{\bibfnamefont{M.}~\bibnamefont{Ding}},
  \bibinfo{journal}{Chaos} \textbf{\bibinfo{volume}{16}}, \bibinfo{eid}{015109}
  (\bibinfo{year}{2006}).

\bibitem[{\citenamefont{Holme}(2015)}]{HolmeEPJB2015}
\bibinfo{author}{\bibfnamefont{P.}~\bibnamefont{Holme}}, \bibinfo{journal}{The
  European Physical Journal B} \textbf{\bibinfo{volume}{88}},
  \bibinfo{pages}{1} (\bibinfo{year}{2015}).

\bibitem[{\citenamefont{Buonomano and
  Merzenich}(1998)}]{BuonomanoMerzenich1998}
\bibinfo{author}{\bibfnamefont{D.~V.} \bibnamefont{Buonomano}}
  \bibnamefont{and} \bibinfo{author}{\bibfnamefont{M.~M.}
  \bibnamefont{Merzenich}}, \bibinfo{journal}{Annual Review of Neuroscience}
  \textbf{\bibinfo{volume}{21}}, \bibinfo{pages}{149} (\bibinfo{year}{1998}),
  \bibinfo{note}{pMID: 9530495}.

\bibitem[{\citenamefont{Belykh et~al.}(2004)\citenamefont{Belykh, Belykh, and
  Hasler}}]{BelykhHasler2004}
\bibinfo{author}{\bibfnamefont{I.~V.} \bibnamefont{Belykh}},
  \bibinfo{author}{\bibfnamefont{V.~N.} \bibnamefont{Belykh}},
  \bibnamefont{and} \bibinfo{author}{\bibfnamefont{M.}~\bibnamefont{Hasler}},
  \bibinfo{journal}{Physica D: Nonlinear Phenomena}
  \textbf{\bibinfo{volume}{195}}, \bibinfo{pages}{188 } (\bibinfo{year}{2004}).

\bibitem[{\citenamefont{Peruani et~al.}(2010)\citenamefont{Peruani, Nicola, and
  Morelli}}]{PeruaniMorelli2010}
\bibinfo{author}{\bibfnamefont{F.}~\bibnamefont{Peruani}},
  \bibinfo{author}{\bibfnamefont{E.~M.} \bibnamefont{Nicola}},
  \bibnamefont{and} \bibinfo{author}{\bibfnamefont{L.~G.}
  \bibnamefont{Morelli}}, \bibinfo{journal}{New Journal of Physics}
  \textbf{\bibinfo{volume}{12}}, \bibinfo{pages}{093029}
  (\bibinfo{year}{2010}).

\bibitem[{\citenamefont{Fujiwara et~al.}(2011)\citenamefont{Fujiwara, Kurths,
  and D\'{\i}az-Guilera}}]{FujiwaraGuilera2011}
\bibinfo{author}{\bibfnamefont{N.}~\bibnamefont{Fujiwara}},
  \bibinfo{author}{\bibfnamefont{J.}~\bibnamefont{Kurths}}, \bibnamefont{and}
  \bibinfo{author}{\bibfnamefont{A.}~\bibnamefont{D\'{\i}az-Guilera}},
  \bibinfo{journal}{Phys. Rev. E} \textbf{\bibinfo{volume}{83}},
  \bibinfo{pages}{025101} (\bibinfo{year}{2011}).

\bibitem[{\citenamefont{Frasca et~al.}(2008)\citenamefont{Frasca, Buscarino,
  Rizzo, Fortuna, and Boccaletti}}]{FrascaBoccaletti2008}
\bibinfo{author}{\bibfnamefont{M.}~\bibnamefont{Frasca}},
  \bibinfo{author}{\bibfnamefont{A.}~\bibnamefont{Buscarino}},
  \bibinfo{author}{\bibfnamefont{A.}~\bibnamefont{Rizzo}},
  \bibinfo{author}{\bibfnamefont{L.}~\bibnamefont{Fortuna}}, \bibnamefont{and}
  \bibinfo{author}{\bibfnamefont{S.}~\bibnamefont{Boccaletti}},
  \bibinfo{journal}{Phys. Rev. Lett.} \textbf{\bibinfo{volume}{100}},
  \bibinfo{pages}{044102} (\bibinfo{year}{2008}).

\bibitem[{\citenamefont{Fujiwara et~al.}(2016)\citenamefont{Fujiwara, Kurths,
  and Díaz-Guilera}}]{FujiwaraGuilera2016}
\bibinfo{author}{\bibfnamefont{N.}~\bibnamefont{Fujiwara}},
  \bibinfo{author}{\bibfnamefont{J.}~\bibnamefont{Kurths}}, \bibnamefont{and}
  \bibinfo{author}{\bibfnamefont{A.}~\bibnamefont{Díaz-Guilera}},
  \bibinfo{journal}{Chaos} \textbf{\bibinfo{volume}{26}}, \bibinfo{eid}{094824}
  (\bibinfo{year}{2016}).

\bibitem[{\citenamefont{Uriu et~al.}(2013)\citenamefont{Uriu, Ares, Oates, and
  Morelli}}]{UriuMorelli2013}
\bibinfo{author}{\bibfnamefont{K.}~\bibnamefont{Uriu}},
  \bibinfo{author}{\bibfnamefont{S.}~\bibnamefont{Ares}},
  \bibinfo{author}{\bibfnamefont{A.~C.} \bibnamefont{Oates}}, \bibnamefont{and}
  \bibinfo{author}{\bibfnamefont{L.~G.} \bibnamefont{Morelli}},
  \bibinfo{journal}{Phys. Rev. E} \textbf{\bibinfo{volume}{87}},
  \bibinfo{pages}{032911} (\bibinfo{year}{2013}).

\bibitem[{\citenamefont{Uriu and Morelli}(2014)}]{UriuMorelli2014}
\bibinfo{author}{\bibfnamefont{K.}~\bibnamefont{Uriu}} \bibnamefont{and}
  \bibinfo{author}{\bibfnamefont{L.~G.} \bibnamefont{Morelli}},
  \bibinfo{journal}{Biophys J} \textbf{\bibinfo{volume}{107}},
  \bibinfo{pages}{514} (\bibinfo{year}{2014}).

\bibitem[{\citenamefont{Nag and Poria}(2016)}]{NagPoriaCS&F2016}
\bibinfo{author}{\bibfnamefont{M.}~\bibnamefont{Nag}} \bibnamefont{and}
  \bibinfo{author}{\bibfnamefont{S.}~\bibnamefont{Poria}},
  \bibinfo{journal}{Chaos, Solitons \& Fractals} \textbf{\bibinfo{volume}{91}},
  \bibinfo{pages}{9 } (\bibinfo{year}{2016}).

\bibitem[{\citenamefont{Stilwell et~al.}(2006)\citenamefont{Stilwell, Bollt,
  and Roberson}}]{StilwayRobertson2006}
\bibinfo{author}{\bibfnamefont{D.~J.} \bibnamefont{Stilwell}},
  \bibinfo{author}{\bibfnamefont{E.~M.} \bibnamefont{Bollt}}, \bibnamefont{and}
  \bibinfo{author}{\bibfnamefont{D.~G.} \bibnamefont{Roberson}},
  \bibinfo{journal}{SIAM Journal on Applied Dynamical Systems}
  \textbf{\bibinfo{volume}{5}}, \bibinfo{pages}{140} (\bibinfo{year}{2006}).

\bibitem[{\citenamefont{Golub and Van~Loan}(1996)}]{GolubVanLoan1996}
\bibinfo{author}{\bibfnamefont{G.~H.} \bibnamefont{Golub}} \bibnamefont{and}
  \bibinfo{author}{\bibfnamefont{C.~F.} \bibnamefont{Van~Loan}},
  \emph{\bibinfo{title}{Matrix Computations}} (\bibinfo{publisher}{Johns
  Hopkins University Press}, \bibinfo{year}{1996}).

\bibitem[{\citenamefont{D'Huys et~al.}(2013)\citenamefont{D'Huys, Zeeb,
  Jüngling, Yanchuk, and Kinzel}}]{DHuysKinzel2013}
\bibinfo{author}{\bibfnamefont{O.}~\bibnamefont{D'Huys}},
  \bibinfo{author}{\bibfnamefont{S.}~\bibnamefont{Zeeb}},
  \bibinfo{author}{\bibfnamefont{T.}~\bibnamefont{Jüngling}},
  \bibinfo{author}{\bibfnamefont{S.}~\bibnamefont{Yanchuk}}, \bibnamefont{and}
  \bibinfo{author}{\bibfnamefont{W.}~\bibnamefont{Kinzel}},
  \bibinfo{journal}{EPL (Europhysics Letters)} \textbf{\bibinfo{volume}{103}}, 
  \bibinfo{pages}{10013},
  (\bibinfo{year}{2013}).

\bibitem[{\citenamefont{J\"ungling et~al.}(2015)\citenamefont{J\"ungling,
  D'Huys, and Kinzel}}]{Jungling2015}
\bibinfo{author}{\bibfnamefont{T.}~\bibnamefont{J\"ungling}},
  \bibinfo{author}{\bibfnamefont{O.}~\bibnamefont{D'Huys}}, \bibnamefont{and}
  \bibinfo{author}{\bibfnamefont{W.}~\bibnamefont{Kinzel}},
  \bibinfo{journal}{Phys. Rev. E} \textbf{\bibinfo{volume}{91}},
  \bibinfo{pages}{062918} (\bibinfo{year}{2015}).

\bibitem[{\citenamefont{Barahona and Pecora}(2002)}]{BarahonaPecora2002}
\bibinfo{author}{\bibfnamefont{M.}~\bibnamefont{Barahona}} \bibnamefont{and}
  \bibinfo{author}{\bibfnamefont{L.~M.} \bibnamefont{Pecora}},
  \bibinfo{journal}{Phys. Rev. Lett.} \textbf{\bibinfo{volume}{89}},
  \bibinfo{pages}{054101} (\bibinfo{year}{2002}).

\bibitem[{\citenamefont{Aviad et~al.}(2012)\citenamefont{Aviad, Reidler,
  Zigzag, Rosenbluh, and Kanter}}]{Aviad2012}
\bibinfo{author}{\bibfnamefont{Y.}~\bibnamefont{Aviad}},
  \bibinfo{author}{\bibfnamefont{I.}~\bibnamefont{Reidler}},
  \bibinfo{author}{\bibfnamefont{M.}~\bibnamefont{Zigzag}},
  \bibinfo{author}{\bibfnamefont{M.}~\bibnamefont{Rosenbluh}},
  \bibnamefont{and} \bibinfo{author}{\bibfnamefont{I.}~\bibnamefont{Kanter}},
  \bibinfo{journal}{Opt. Express} \textbf{\bibinfo{volume}{20}},
  \bibinfo{pages}{4352} (\bibinfo{year}{2012}).

\bibitem[{\citenamefont{Grabow et~al.}(2015)\citenamefont{Grabow, Grosskinsky,
  Kurths, and Timme}}]{GrabowJurgenPRE2015}
\bibinfo{author}{\bibfnamefont{C.}~\bibnamefont{Grabow}},
  \bibinfo{author}{\bibfnamefont{S.}~\bibnamefont{Grosskinsky}},
  \bibinfo{author}{\bibfnamefont{J.}~\bibnamefont{Kurths}}, \bibnamefont{and}
  \bibinfo{author}{\bibfnamefont{M.}~\bibnamefont{Timme}},
  \bibinfo{journal}{Phys. Rev. E} \textbf{\bibinfo{volume}{91}},
  \bibinfo{pages}{052815} (\bibinfo{year}{2015}).

\bibitem[{\citenamefont{Newman and Watts}(1999)}]{Newman1999}
\bibinfo{author}{\bibfnamefont{M.}~\bibnamefont{Newman}} \bibnamefont{and}
  \bibinfo{author}{\bibfnamefont{D.}~\bibnamefont{Watts}},
  \bibinfo{journal}{Physics Letters A} \textbf{\bibinfo{volume}{263}},
  \bibinfo{pages}{341 } (\bibinfo{year}{1999}).

\bibitem[{\citenamefont{Watts and Strogatz}(1998)}]{WattsStrogatz1998}
\bibinfo{author}{\bibfnamefont{D.~J.} \bibnamefont{Watts}} \bibnamefont{and}
  \bibinfo{author}{\bibfnamefont{S.~H.} \bibnamefont{Strogatz}},
  \bibinfo{journal}{Nature} \textbf{\bibinfo{volume}{393}},
  \bibinfo{pages}{409} (\bibinfo{year}{1998}).

\bibitem[{\citenamefont{Kühn}(2008)}]{KuhnJPA2008}
\bibinfo{author}{\bibfnamefont{R.}~\bibnamefont{Kühn}},
  \bibinfo{journal}{Journal of Physics A: Mathematical and Theoretical}
  \textbf{\bibinfo{volume}{41}}, \bibinfo{pages}{295002}
  (\bibinfo{year}{2008}).

\bibitem[{\citenamefont{Billen et~al.}(2009)\citenamefont{Billen, Wilson,
  Baljon, and Rabinovitch}}]{BillenRabinovitchPRE2009}
\bibinfo{author}{\bibfnamefont{J.}~\bibnamefont{Billen}},
  \bibinfo{author}{\bibfnamefont{M.}~\bibnamefont{Wilson}},
  \bibinfo{author}{\bibfnamefont{A.}~\bibnamefont{Baljon}}, \bibnamefont{and}
  \bibinfo{author}{\bibfnamefont{A.}~\bibnamefont{Rabinovitch}},
  \bibinfo{journal}{Phys. Rev. E} \textbf{\bibinfo{volume}{80}},
  \bibinfo{pages}{046116} (\bibinfo{year}{2009}).

\bibitem[{\citenamefont{Knoblauch et~al.}(2012)\citenamefont{Knoblauch, Hauser,
  Gewaltig, K{\"o}rner, and Palm}}]{knoblauch2012}
\bibinfo{author}{\bibfnamefont{A.}~\bibnamefont{Knoblauch}},
  \bibinfo{author}{\bibfnamefont{F.}~\bibnamefont{Hauser}},
  \bibinfo{author}{\bibfnamefont{M.-O.} \bibnamefont{Gewaltig}},
  \bibinfo{author}{\bibfnamefont{E.}~\bibnamefont{K{\"o}rner}},
  \bibnamefont{and} \bibinfo{author}{\bibfnamefont{G.}~\bibnamefont{Palm}},
  \bibinfo{journal}{Frontiers in computational neuroscience}
  \textbf{\bibinfo{volume}{6}}, \bibinfo{pages}{55} (\bibinfo{year}{2012}).

\bibitem[{\citenamefont{Harmer and Abbott}(1999)}]{Harmer1999}
\bibinfo{author}{\bibfnamefont{G.~P.} \bibnamefont{Harmer}} \bibnamefont{and}
  \bibinfo{author}{\bibfnamefont{D.}~\bibnamefont{Abbott}},
  \bibinfo{journal}{Nature} \textbf{\bibinfo{volume}{402}},
  \bibinfo{pages}{864} (\bibinfo{year}{1999}).

\bibitem[{\citenamefont{Landau and Lifshitz}(1960)}]{LandauMechanics}
\bibinfo{author}{\bibfnamefont{L.}~\bibnamefont{Landau}} \bibnamefont{and}
  \bibinfo{author}{\bibfnamefont{E.}~\bibnamefont{Lifshitz}},
  \emph{\bibinfo{title}{Mechanics}} (\bibinfo{publisher}{Pergamon Press},
  \bibinfo{year}{1960}).

\bibitem[{\citenamefont{Elaydi}(2007)}]{elaydi2007discrete}
\bibinfo{author}{\bibfnamefont{S.~N.} \bibnamefont{Elaydi}},
  \emph{\bibinfo{title}{Discrete chaos: with applications in science and
  engineering}} (\bibinfo{publisher}{CRC Press}, \bibinfo{year}{2007}).

\end{thebibliography}

\end{document}